\documentclass[iop]{emulateapj}
\usepackage{multirow}
\usepackage{booktabs}
\usepackage[figuresright]{rotating}
\usepackage{subfigure}
\usepackage{epic,eepic}
\usepackage{graphicx}
\usepackage{graphics}
\usepackage{hyperref}
\usepackage{threeparttable}

\shorttitle{STELLAR X-RAY ACTIVITY}
\shortauthors{He et al.}

\begin{document}
\title{A COMBINED {\it CHANDRA} and LAMOST STUDY OF STELLAR ACTIVITY}

\author{Lin He\altaffilmark{1,2}, Song Wang\altaffilmark{2}, Jifeng Liu\altaffilmark{2,3}, Roberto Soria\altaffilmark{3,4,5}, Zhongrui Bai\altaffilmark{2}, Huiqin Yang\altaffilmark{2,3}, Yu Bai\altaffilmark{2}, Jincheng Guo\altaffilmark{6}}

\altaffiltext{1}{School of Astronomy and Space Science and Key Laboratory of Modern Astronomy and Astrophysics, Nanjing University, Nanjing, 210093, P. R. China}
\altaffiltext{2}{Key Laboratory of Optical Astronomy, National Astronomical Observatories,
Chinese Academy of Sciences, Beijing 100101, China; songw@bao.ac.cn}
\altaffiltext{3}{College of Astronomy and Space Sciences,
University of Chinese Academy of Sciences, Beijing 100049, China}
\altaffiltext{4}{International Centre for Radio Astronomy Research, Curtin University, GPO Box U1987, Perth, WA 6845, Australia}
\altaffiltext{5}{Sydney Institute for Astronomy, School of Physics A28, The University of Sydney, Sydney, NSW 2006, Australia}
\altaffiltext{6}{Department of Astronomy, Peking University, Beijing 100871, China}

\begin{abstract}
We probed stellar X-ray activity over a wide range of stellar parameters, using {\it Chandra} and LAMOST data.
We measured the X-ray-to-bolometric luminosity ratio ($R_X = L_X/L_{\rm bol}$) for 484 main-sequence stars,
and found a bimodal distribution for G and K types.
We interpret this bimodality as evidence of two sub-populations with different coronal temperatures,
which are caused by different coronal heating rates.
Using the metallicity and velocity information,
we find both of the two sub-populations are mostly located in the thin disk.
We find no trend of $R_X$ with stellar age for stars older than $\sim$ 4 Gyr;
there is a trough in the $R_X$ vs age distribution, with the lowest range of $R_X$ appearing at ages around 2 Gyr.
We then examined the correlation between $R_X$ and $R_{\rm H\alpha}$ (proxy of chromospheric activity):
we find that the two quantities are well correlated, as found in many earlier studies.
Finally, we selected a sample of twelve stars with X-ray flares,
and studied the light-curve morphology of the flares.
The variety of flare profiles and timescales observed in our sample
suggests the contribution of different processes of energy release.
\end{abstract}


\section{INTRODUCTION}
\label{intro.sec}

Stellar activity has been extensively explored in the X-ray regime for nearly forty years
\citep{Vaiana1981, Schmitt1985, Rosner1985}.
From the $Einstein$ mission, X-ray observations have shown that
X-ray emission is prevalent among almost all stellar classes
\citep[e.g.,][]{Stocke1983, Maccacaro1988, Stocke1991},
although with different mechanisms for early and late-type stars \citep[see][for a review]{Narain1990, Gudel2004}.
In general, the hottest and most massive stars emit X-rays arising from either small-scale shocks in their winds
or collisions between the wind and circumstellar material \citep{Lucy1980, Parkin2009},
while X-ray emission from late-type stars is attributed to the presence of a magnetic corona \citep{Vaiana1981}.
Previous studies have made significant progress, both observationally and
theoretically, in understanding the correlation between corona and magnetic fields \citep[e.g.,][]{Klimchuk2006,Cranmer2007}.
However, some detailed physical processes, such as coronal heating,
are still poorly understood, and remain open issues in current astrophysics.

The study of stellar activity is important because it can provide new constraints for magnetic dynamo models.
A number of proxies have been used to study the activity of different stellar components,
such as the corona (X-ray emission),
the chromosphere (Ca II HK; Mg II; H$\alpha$; optical flares; NUV flux),
the transition region (FUV flux),
and the photosphere (star spots).
Observations have revealed empirical scalings of these proxies, and therefore of magnetic activity,
with rotation period or Rossby number,
the latter of which is a ratio of rotation period to convective turnover time
\citep[e.g., ][]{Pallavicini1981, Vilhu1984, Micela1985, Pizzolato2003, Wright2011, Gondoin2012, Mathur2014, Blackman2015}.
It is believed that the generation of magnetic energy by large-scale dynamo action
is driven by rotation and convection \citep{Charbonneau2010, Reiners2014}.
In the solar-type dynamo mechanism ($\alpha\Omega$ dynamo), the magnetic field is generated
in the deep convection zones due to the interior radial differential rotation;
the magnetic field then rises to the photosphere and produces chromospheric heating
through the interaction with the uppermost convection zones \citep[e.g.,][]{Cardini2007,Cincunegui2007}.

In parallel with the uncertainty on the physical heating processes,
the magnetic activity evolution also remains unclear.
Although many authors have investigated the dependence of chromospheric and coronal emission levels on
the stellar age and evolutionary stage,
the results are still under debate and improved \citep{Pizzolato2003,Lyra2005,Pace2013,Zhao2013,Booth2017}.
In general, stellar activity and rotation are observed to decay with stellar age
\citep{Wilson1963, Kraft1967, Skumanich1972, Simon1985, Pace2004, Cardini2007,Mamajek2008}.
Furthermore, the coronal and chromospheric emission have been used as age indicators
\citep{Soderblom1991, Donahue1993, Lachaume1999, Mamajek2008, Soderblom2010, Jeffries2014}.
However, the age-activity relation constructed from X-ray emission is quite uncertain, since the X-ray flux is
significantly variable \citep{Soderblom2010}.
More effort is needed to better understand the activity evolution and
to improve the age-activity relation as a dating technique.

A large sample of late-type stars with known stellar parameters
is well suitable to investigate stellar activity and to
explore the correlation between stellar activity and other stellar properties.
These are helpful for understanding coronal heating processes, as well as the evolution of stellar X-ray activity.
Moreover, the study of magnetic activity, especially the UV and X-ray emission,
helps us refine our knowledge of the habitable zone of exoplanets \citep[e.g.,][]{Gudel2014,Robertson2015,Gallet2017}.
In this paper, we use the $Chandra$ data archive and the Large Sky Area Multi-Object Fiber Spectroscopic Telescope (LAMOST) database
to provide a sample of stars with X-ray emission and accurate stellar parameters.
In Section \ref{analysis.sec}, we introduce the sample selection and data reduction process.
Section \ref{result.sec} presents the results and some discussions,
including the X-ray activity in different stellar types,
the bimodality of X-ray activity of G and K types,
the relation between X-ray activity and age,
the relation between X-ray and chromospheric activity,
and the light-curve morphology of flares.

\section{SAMPLE SELECTION AND DATA REDUCTION}
\label{analysis.sec}

\subsection{Sample Selection}
\label{sample.sec}

To estimate the $R_X$ ($= L_X/L_{\rm bol}$) of different stellar types,
we first cross-matched the $Chandra$ point source catalog \citep{Wang2016b} and the LAMOST database (DR4),
using a radius of 3$^{\prime\prime}$.
This led to 2638 unique $Chandra$ sources with 3502 LAMOST spectral observations.
To calculate the likelihood of mismatch,
we shifted the positions of {\it Chandra} sources by 1$^{\prime}$,
and cross-matched them with the LAMOST catalog again using the same radius.
In this case, we obtained 76 matches,
and we conclude the likelihood of mismatch is about 2.88$\%$.

In order to have a reliable estimation of stellar parameters, we only used spectra with
signal-to-noise ratio (SNR) higher than 7 in the $r$ band.
There are four kinds of classes in the LAMOST database, including ``STAR'', ``GALAXY'', ``QSO'', and ``Unknown'',
and we excluded those objects with the latter three classifications.
In addition, some objects classified as ``STAR'' are actually globular clusters or galaxies.
To exclude them, we cross-matched the sample sources with SIMBAD database
\footnote{http://simbad.u-strasbg.fr/simbad/}
and checked the spectra by eye.
Young stellar objects were also excluded following SIMBAD classification
(e.g., young stellar object, pre-main sequence star, Orion type, T Tau star, Herbig-Haro Object, and Herbig Ae/Be star).
In summary, up to 2023 LAMOST observations were excluded, including
582 observations with low SNR,
311 observations classified as ``Unknown'',
41 observations of 31 globular clusters,
757 observations of 648 galaxies and QSOs,
254 observations of 179 young stellar objects,
71 observations of 31 binaries,
and seven observations of four other sources
(one planetary nebula, two white dwarfs, and one cataclysmic variable).
This led to 1086 unique X-ray sources with 1479 spectral observations.

\subsection{$Chandra$ Data Analysis}
\label{chandra.sec}

The subarcsecond spatial resolution of {\it Chandra} allows X-ray sources
 to be matched unambiguously to multiwavelength counterparts.
Our primary database is the $Chandra$ point source catalog \citep{Wang2016b},
which includes 363,530 source detections, belonging to 217,828 distinct X-ray sources.
To determine the photon counts of the sample sources,
firstly we used {\tt wavdetect} to construct the $3\sigma$ elliptical source region for each detection,
by fitting a 2D elliptical Gaussian to the distribution of (observed) counts in the source cell.
The broad band count rate (0.5--7 keV) and flux were then extracted with {\tt srcflux},
which determines the net count rate (NET\_RATE\_APER) and model-independent flux (NET\_FLUX\_APER) given source and background regions.
We defined the background region as an elliptical annulus around the source region,
with the inner/outer radii as 2/4 times the radii of the source ellipse.

Some sources in our sample were observed more than once by $Chandra$.
In some cases, a source was detected by {\tt wavdetect} in one observation
but not detected in another observation.
We calculate an upper limit using the 4$\sigma$ detection threshold,
which is a function of the off-axis angle and the background level
\citep[see][for details]{Wang2016a}.

\subsection{LAMOST Data}
\label{lamost.sec}

The LAMOST (also called the Guo Shou Jing Telescope) is a reflecting Schmidt telescope
with an effective aperture of 4 m and a field of view of 5 degrees \citep{Cui2012, Zhao2012}.
Four thousand fibres are configured on the focal surface,
which enables the observation of up to 4000 objects simultaneously.
The spectral resolution is $R \sim$ 1800 over the wavelength range of 3690--9100 {\AA}.
After a five-year regular survey \footnote{http://www.lamost.org/public/node/311?locale=en}
started in 2012, more than eight million spectra
have been obtained.
Stellar parameters, including effective temperature,
surface gravity, metallicity, radial velocity, distance, and reddening,
have been estimated by several studies \citep[e.g.,][]{Wu2011, Wang2016, Xiang2017b}.
Such parameters can be used in many fields and may eventually revolutionize our knowledge
of stellar evolution and the structure of the Milky Way.
In this paper, we used the spectra from the DR4 database.

\subsection{$Photometric$ Data}
\label{photometric.sec}

\begin{figure}[htb]
\center
\includegraphics[width=0.49\textwidth]{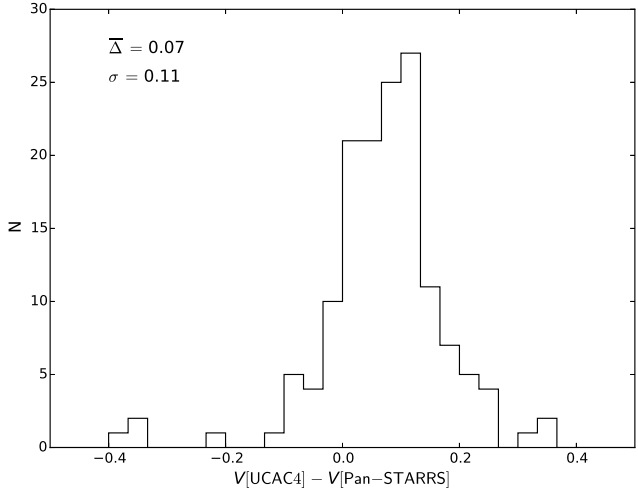}
\caption[]{Comparison of $V$ magnitudes from UCAC4 and Pan-STARRS for common objects in the sample.}
\label{vband.fig}
\end{figure}

We first collected the $V$-band magnitude from the UCAC4 catalog \citep{Zacharias2013}.
For objects without a UCAC4 $V$-band magnitude,
we calculated it using the $g$ and $r$ magnitudes from Pan-STARRS \citep{Flewelling2016}, following \citet{Jester2005}:
\begin{equation}
V = g - 0.59 \times (g - r) - 0.01.
\end{equation}
There is a small systematic offset ($\sim$ 0.1 mag) between the $V$ magnitudes of UCAC4 and Pan-STARRS,
estimated using common objects in the sample (Figure \ref{vband.fig}).

\begin{figure}[htb]
\center
\includegraphics[width=0.49\textwidth]{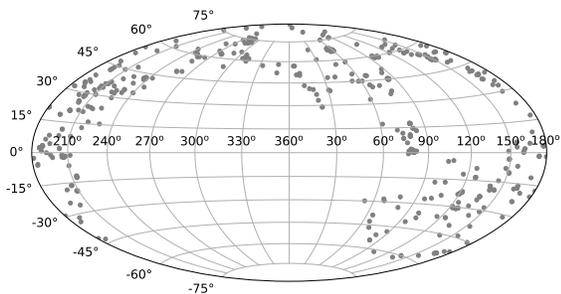}
\caption[]{Sky distribution of the sample sources in this paper, in Galactic coordinates.}
\label{gal.fig}
\end{figure}

In summary, there are 484 X-ray sources with complete stellar parameters
(i.e., $V$ mag, reddening $E(B-V)$, effective temperature $T_{\rm eff}$, surface gravity log$g$, and distance).
We listed X-ray properties in each ${\it Chandra}$ observation for the sample sources in Table \ref{xray.tab}.
We also listed LAMOST parameters for the sample stars in Table \ref{lamost.tab},
including the subclass classification from \citet{Luo2015};
the $T_{\rm eff}$, log$g$, $E(B-V)$,
[Fe/H], [$\alpha$/Fe],
and distance from \citet{Xiang2017b};
the age and mass from \citet{Xiang2017a}.
Figure \ref{gal.fig} shows the sky distribution of the sample sources in Galactic coordinates.

\section{Results and Discussion}
\label{result.sec}

In this paper, we used the ratio of X-ray to bolometric luminosities, $R_X$, as the X-ray activity indicator.
%
In the X-ray band, first, we calculated an exposure-weighted averaged X-ray flux for each star.
Then, we converted it to the 0.3--8 keV band using PIMMS
\footnote{http://cxc.harvard.edu/toolkit/pimms.jsp}
 with an APEC model \citep[coronal temperature log$T =$ 6.5:][]{Schmitt1990}.
The hydrogen column density ($N_{\rm H}$) was converted from individual optical extinction:
\begin{equation}
N_{\rm H} {\rm (cm ^{-2})} = 2.19 \times 10^{21} A_V {\rm (mag)}
\label{nh.eq}
\end{equation}
\citep{Zhu2017}.
The X-ray luminosity ($L_X$) was determined from the unabsorbed flux ($f_X$) and distance.
The bolometric magnitude ($M_{\rm bol}$) was estimated from the
apparent $V$-band magnitude, reddening, bolometric correction, and distance:
\begin{equation}
M_{\rm bol} = V - A_V - BC_{V} - DM,
\label{mbol.eq}
\end{equation}
where $A_V=3.1\times~E(B-V)$,
$BC_{V}$ were obtained from \citet{Bertone2004} for dwarfs and giants with different temperatures,
and $DM$ is the distance modulus term calculated from the distance.
The bolometric luminosity ($L_{\rm bol}$) was then calculated using
\begin{equation}
{\rm log}~L_{\rm bol} = -0.4\times~(M_{\rm bol}-M_{\rm bol, \odot}) + {\rm log}~L_{\rm bol, \odot}.
\label{lx.eq}
\end{equation}
The bolometric magnitude and luminosity for the Sun is 4.74 mag and $3.84\times10^{33}$ erg s$^{-1}$ \citep{Carroll1996}.
Finally, the $R_X$ was calculated for the sample stars.

We should note that the unabsorbed X-ray flux, converted from the absorbed X-ray flux using PIMMS, 
is dependent on the coronal temperature set in the APEC model.
To evaluate the influence on the $R_X$, we re-estimated the unabsorbed X-ray flux with a higher temperature (log$T = 7$).
Using the median value of $N_{\rm H}$ ($5.2\times10^{20}$ cm$^{-2}$) for the sample stars, the unabsorbed X-ray flux $f_X$ decreases by $\sim$ 28\%. This will lead to a decrease of log$R_X$ by $\sim$ 0.14.
The variation of $R_X$ would not affect the following analysis too much.

We also calculated the X-ray hardness ratio, $HR = (c2-c1)/(c1+c2)$,
where $c1$ is the counts of the soft band (0.3--1 keV) and
$c2$ is the counts of the hard band (1--8 keV).
These parameters are listed in Table \ref{lx2lbol.tab}.

\begin{table*}
\begin{center}
\scriptsize
\caption[]{X-ray information for the stars in our sample.}
\label{xray.tab}
\begin{tabular}{ccccccccc}
\hline\noalign{\smallskip}
Object  &   ObsID &   Detector   &    Exp. Time   &   MJD   &  VigF  &  Count Rate  &  absorbed flux  & Flag\\
       &          &               &      (s)      &         &        &   (ks$^{-1}$)  & (10$^{-15}$ erg cm$^{-2}$ s$^{-1}$)   &   \\
  (1)  &     (2) &       (3)      &       (4)     &   (5)   &   (6)  &  (7)         &  (8)                 & (9) \\
\hline\noalign{\smallskip}
J001208.1+503015 & 1864 & s3 & 5876.3 & 51898 & 0.947 & 1.53$^{-0.75}_{+1.06}$ & 3.3$^{-1.6}_{+2.3}$ & \\
J001313.2+000250 & 4829 & s3 & 6658.7 & 53183 & 0.975 & 0.58$^{-0.37}_{+0.66}$ & 4.2$^{-2.8}_{+4.9}$ & \\
J002756.6+261651 & 14012 & i3 & 21692.2 & 56194 & 0.762 & 16.0$^{-1.4}_{+1.4}$ & 184$^{-16}_{+16}$ & \\
   & 3249 & i3 & 9977.0 & 52451 & 0.975 & 34.7$^{-3.1}_{+3.1}$ & 259$^{-23}_{+23}$ & \\
J003254.4+393821 & 12991 & s2 & 34587.1 & 55738 & 0.636 & 0.18$^{-0.1}_{+0.16}$ & 3.2$^{-1.8}_{+2.7}$ & \\
   & 9525 & s3 & 34732.2 & 54739 & 0.841 & 0.27$^{-0.13}_{+0.19}$ & 1.61$^{-0.80}_{+1.11}$ & \\
J003802.9+400824 & 2046 & s2 & 14765.9 & 51853 & 0.579 & 1.36$^{-0.53}_{+0.65}$ & 18.8$^{-7.3}_{+9.0}$ & \\
J003823.9+401250 & 2046 & s2 & 14765.9 & 51853 & 0.67 & 28.5$^{-2.3}_{+2.3}$ & 176$^{-14}_{+14}$ & \\
   & 2048 & s3 & 13772.2 & 52093 & 0.513 & 19.6$^{-2.0}_{+2.0}$ & 93.5$^{-9.3}_{+9.4}$ & \\
J003831.2+401711 & 2046 & s3 & 14762.8 & 51853 & 0.994 & 0.8$^{-0.33}_{+0.46}$ & 1.59$^{-0.66}_{+0.91}$ & \\
   & 2047 & s3 & 14585.9 & 51974 & 0.974 & 1.02$^{-0.38}_{+0.5}$ & 2.9$^{-1.1}_{+1.5}$ & \\
   & 2048 & s3 & 13772.2 & 52093 & 0.924 & 0.35$^{-0.21}_{+0.34}$ & 1.01$^{-0.61}_{+1.01}$ & \\
J004118.6+405159 & 2049 & s2 & 14568.4 & 51853 & 0.564 & 8.6$^{-1.3}_{+1.3}$ & 50.4$^{-7.6}_{+7.7}$ & \\
   & 2902 & i0 & 4695.0 & 52614 & 0.798 & 7.3$^{-1.9}_{+2.3}$ & 57$^{-15}_{+18}$ & \\
J004301.5+411052 & 10551 & i2 & 3964.7 & 54840 & 0.805 & $<$ 2.26 & $<$ 29.5 & u\\
\noalign{\smallskip}\hline
\end{tabular}
\end{center}
\tablecomments{The columns are:
(1) object;
(2) observation ID;
(3) the detector on $Chandra$ ACIS;
(3) exposure time after deadtime correction;
(4) Modified Julian Date for the beginning point of the observation;
(6) vignetting factor;
(7) net count rate in the 0.5--7 keV;
(8) absorbed X-ray flux in the 0.5--7 keV.
(9) ``u'' means the source is not detected in this observation, and an upper limit is estimated.}
(This table is available in its entirety in machine-readable and Virtual Observatory (VO) forms in the online journal. A portion
is shown here for guidance regarding its form and content.)
\end{table*}

\begin{table*}
\begin{center}
\scriptsize
\caption{LAMOST information for the stars in our sample.}
\label{lamost.tab}
\begin{tabular}{cccccccccc}
\hline\noalign{\smallskip}
Object  &   Subclass &   $T_{\rm eff}$   &    log$g$   &   E(B-V)   & [Fe/H]  &  [$\alpha$/Fe]   &   Distance  &     Age  &  Mass         \\
        &           &       (K)        &             &            &         &                  &       (pc)  &    (Gyr) &  ($M_{\odot}$)  \\
  (1)   &     (2)   &       (3)        &        (4)  &      (5)   &   (6)   &  (7)             &       (8)   &  (9)     &  (10)          \\
\hline\noalign{\smallskip}
J001208.1+503015 & G2 & 6010$\pm$61 & 4.56$\pm$0.11 & 0.07 & -0.07$\pm$0.06 & 0.04$\pm$0.1 & 265$\pm$26 & ... & ...\\
J001313.2+000250 & F6 & 6340$\pm$117 & 4.3$\pm$0.22 & 0.09 & 0.08$\pm$0.17 & 0.15$\pm$0.14 & 641$\pm$218 & ... & ...\\
J002756.6+261651 & G9 & 5388$\pm$146 & 4.33$\pm$0.11 & 0.04 & -0.05$\pm$0.08 & 0.12$\pm$0.1 & 147$\pm$13 & ... & ...\\
J003254.4+393821 & K1 & 5446$\pm$211 & 4.41$\pm$0.11 & 0.08 & 0.31$\pm$0.11 & ... & 723$\pm$127 & ... & ...\\
J003802.9+400824 & G2 & 5871$\pm$60 & 4.61$\pm$0.11 & 0.05 & 0.02$\pm$0.06 & 0.06$\pm$0.1 & 272$\pm$23 & ... & ...\\
J003823.9+401250 & F9 & 5729$\pm$60 & 3.38$\pm$0.09 & 0.06 & -0.04$\pm$0.1 & 0.04$\pm$0.1 & 1286$\pm$138 & ... & ...\\
J003831.2+401711 & G8 & 5259$\pm$121 & 4.03$\pm$0.18 & 0.1 & -0.01$\pm$0.14 & 0.03$\pm$0.11 & 1337$\pm$364 & ... & ...\\
J004118.6+405159 & G3 & 5732$\pm$68 & 4.36$\pm$0.11 & 0.11 & -0.13$\pm$0.09 & 0.04$\pm$0.1 & 506$\pm$87 & 12.0$\pm$2.9 & 0.91$\pm$0.06\\
\noalign{\smallskip}\hline
\end{tabular}
\end{center}
(This table is available in its entirety in machine-readable and Virtual Observatory (VO) forms in the online journal. A portion
is shown here for guidance regarding its form and content.)
\end{table*}

\begin{table*}
\begin{center}
\scriptsize
\caption[]{Key results for the stars in our sample.}
\label{lx2lbol.tab}
\begin{tabular}{cccccccc}
\hline\noalign{\smallskip}
Object  &     $V$  &   $M_{\rm bol}$  &   unabsorbed $f_X$                 & $L_X$             &     log$R_X$  &  $HR$     &   Flag \\
        &   (mag)  &  (mag)      &  (10$^{-15}$ erg cm$^{-2}$ s$^{-1}$)  &   (erg s$^{-1}$)  &            &           &             \\
  (1)   &     (2)  &    (3)      &         (4)                        &      (5)          &   (6)     &     (7)    & (8)        \\
\hline\noalign{\smallskip}
J001208.1+503015 & 11.71$\pm$0.09 & 4.57 & 6.2$\pm$3.5 & 5.2e+28$\pm$3.2e+28 & -4.94$\pm$0.27 & -0.43$\pm$0.53 & u\\
J001313.2+000250 & 12.48$\pm$0.11 & 3.46 & 8.2$\pm$7.2 & 4.0e+29$\pm$4.5e+29 & -4.49$\pm$0.48 & -0.52$\pm$0.99 & u\\
J002756.6+261651 & 11.62$\pm$0.1 & 5.63 & 350$\pm$32 & 9.0e+29$\pm$1.8e+29 & -3.27$\pm$0.10 & 0.1$\pm$0.04 & u\\
J003254.4+393821 & 14.55$\pm$0.1 & 5.10 & 4.5$\pm$2.2 & 2.8e+29$\pm$1.7e+29 & -3.99$\pm$0.27 & 0.72$\pm$0.34 & u\\
J003802.9+400824 & 12.21$\pm$0.11 & 4.98 & 33$\pm$14 & 2.9e+29$\pm$1.3e+29 & -4.03$\pm$0.20 & -0.22$\pm$0.33 & u\\
J003823.9+401250 & 12.3$\pm$0.11 & 1.71 & 244$\pm$22 & 4.8e+31$\pm$1.1e+31 & -3.11$\pm$0.11 & 0.21$\pm$0.04 & u\\
J003831.2+401711 & 15.4$\pm$0.12 & 4.54 & 3.7$\pm$1.1 & 8.0e+29$\pm$5.0e+29 & -3.76$\pm$0.27 & -0.5$\pm$0.2 & u\\
J004118.6+405159 & 12.69$\pm$0.11 & 4.07 & 109$\pm$16 & 3.3e+30$\pm$1.2e+30 & -3.33$\pm$0.17 & -0.11$\pm$0.09 & u\\
\noalign{\smallskip}\hline
\end{tabular}
\end{center}
\tablecomments{The columns are:
(1) object;
(2) extinction-corrected $V$-band magnitude;
(3) bolometric magnitude;
(4) unabsorbed X-ray flux in the 0.3--8 keV;
(5) X-ray luminosity in the 0.3--8 keV.
(6) X-ray-to-bolometric luminosity ratio;
(7) Hardness ratio $HR = (c2-c1)/(c1+c2)$. $c1$ and $c2$ represent background-subtracted counts in soft (0.3-1 keV) and hard (1-8 keV) band, respectively.
(8) ``u'' means $V$ magnitude from UCAC4, while ``p'' means $V$ magnitude from Pan-STARRS.}
(This table is available in its entirety in machine-readable and Virtual Observatory (VO) forms in the online journal. A portion
is shown here for guidance regarding its form and content.)
\end{table*}

\subsection{X-ray Activity in Different Stellar Types}
\label{types.sec}

\begin{figure*}[htb]
\center
\includegraphics[width=0.49\textwidth]{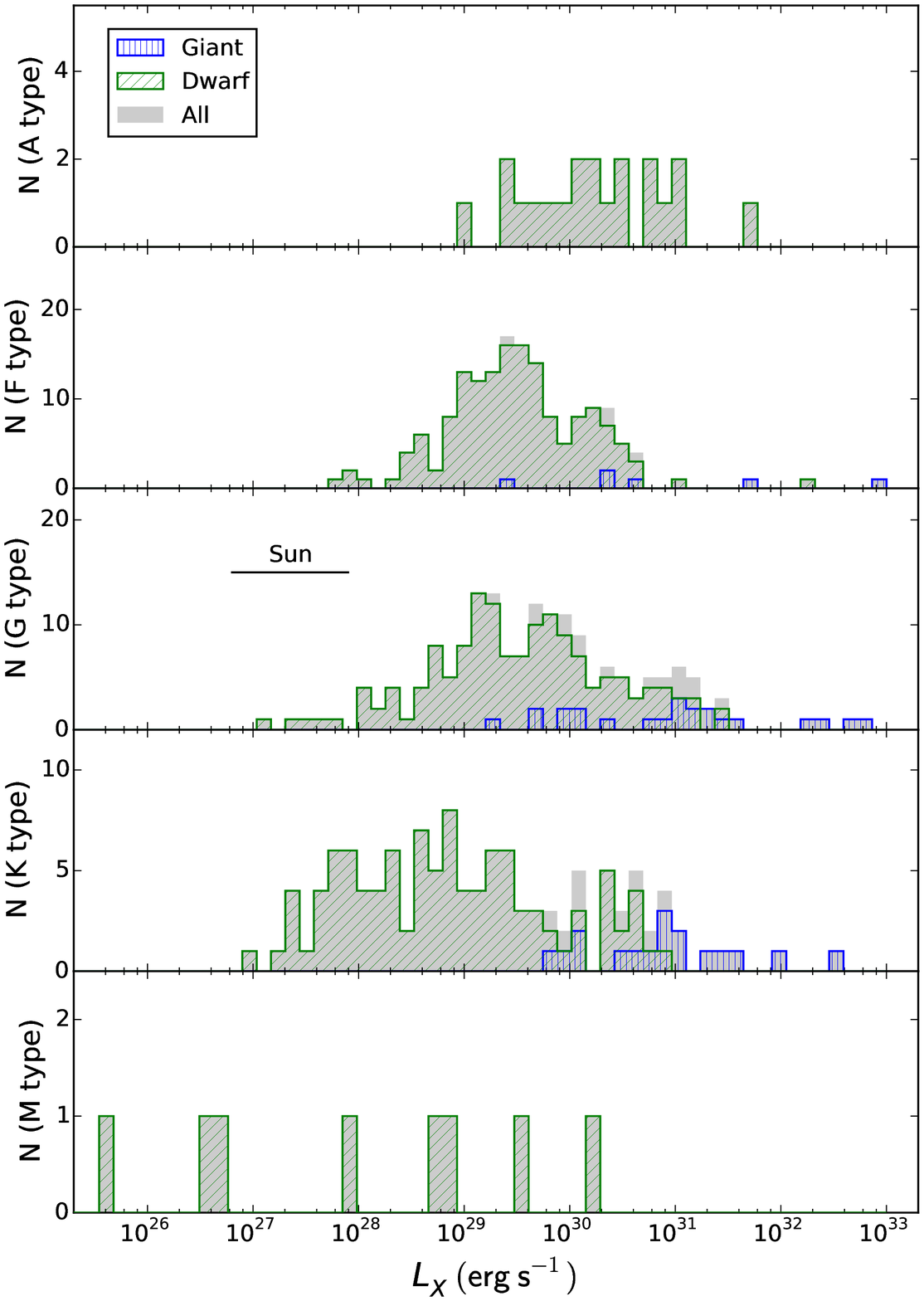}
\includegraphics[width=0.49\textwidth]{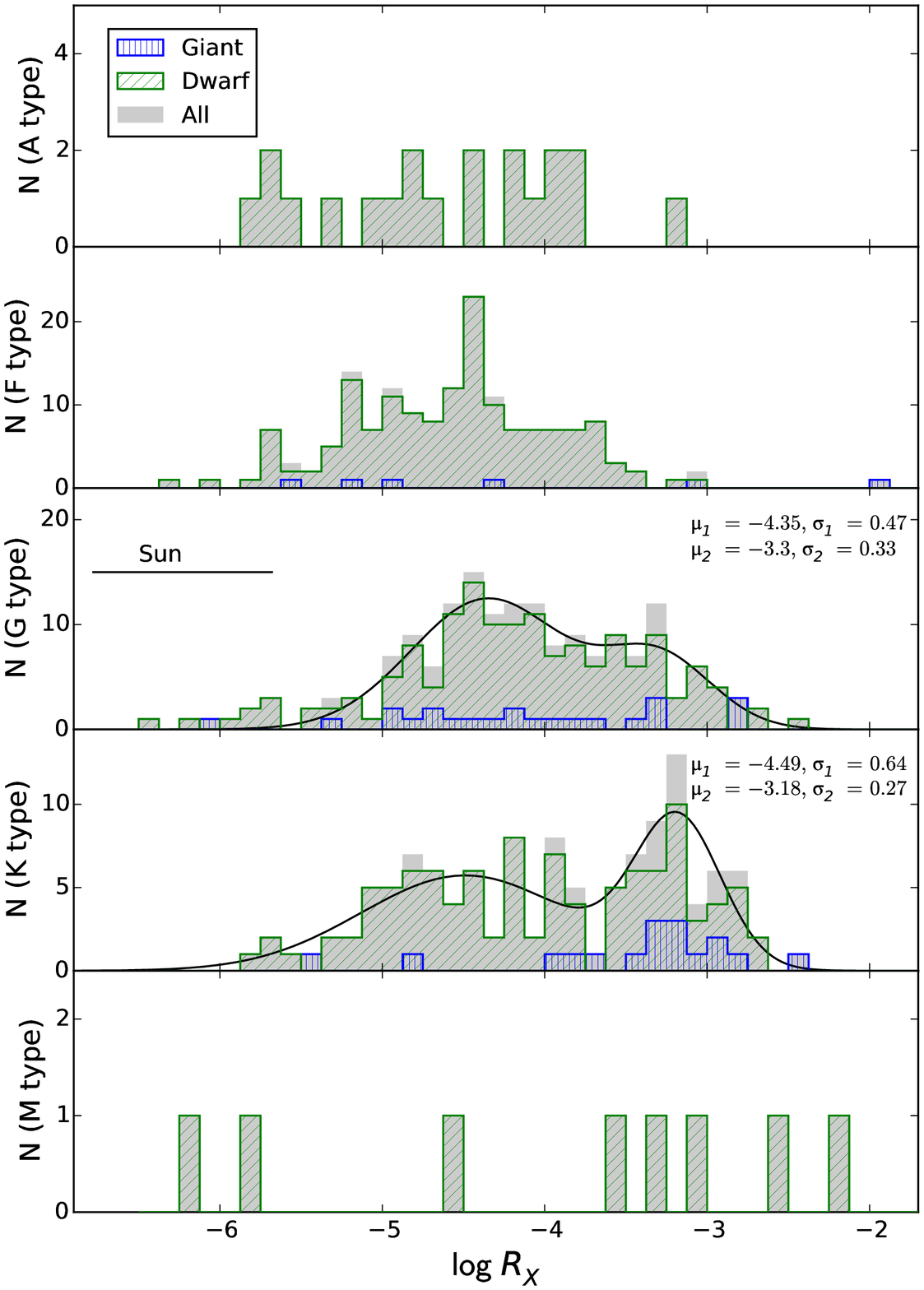}
\caption[]{Left panel: Histograms showing the distributions of $L_X$ for stars from A to M type.
These histograms plotted with blue vertical lines, green slashes, and shaded area indicate giants, dwarfs, and all stars.
The range of solar X-ray luminosity is 10$^{26.8}$--10$^{27.9}$ ergs s$^{-1}$ \citep{Judge2003}, covering a typical solar cycle.
Right panel: Histograms showing the distributions of $R_X$ for stars of spectral type A through M.
These histograms plotted with blue vertical lines, green slashes, and shaded area indicate giants, dwarfs, and all stars.
The log$R_X$ of the Sun varies from $-$6.8 to $-$5.7 in a typical solar cycle.}
\label{subclass.fig}
\end{figure*}

\begin{figure}[htb]
\center
\includegraphics[width=0.48\textwidth]{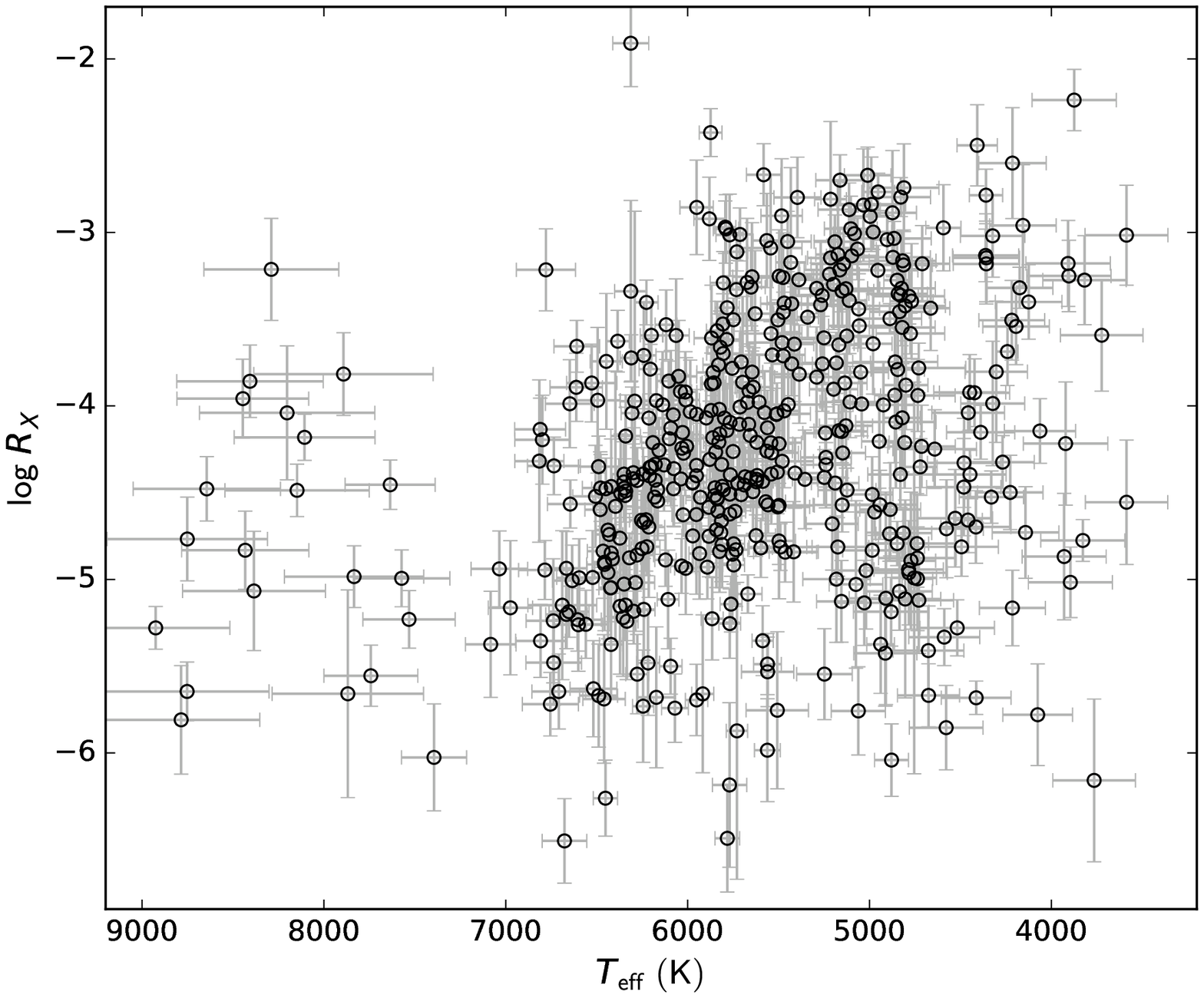}
\caption[]{$R_X$ versus $T_{\rm eff}$.}
\label{lx2lbolT.fig}
\end{figure}

Figure \ref{subclass.fig} displays the distributions of $L_X$
and $R_X$ for stars from spectral type A through M.
Most M type stars in LAMOST have poorly constrained values of log$g$ and distance.
Therefore, only few M type stars were included in our sample.

The $L_X$ distribution of late-type stars (G-type or later) extends to significantly lower luminosities
than early-type stars.
The F, G, and K stars show a fairly wide range of emission levels, with X-ray luminosity ranging from $10^{27}$
to $10^{33}$ ergs s$^{-1}$. In these types, giants are clearly X-ray brighter than dwarfs.
The F, G, and K giants are the brightest X-ray sources in our sample.

In general, late-type stars have more active ones (higher $R_X$) than early-type stars.
Considering the similarity with solar emission, such as thermal emission-line X-ray spectrum and flares,
it was suggested that X-ray emission from late-type stars originates from stellar coronae
that are similar to the Sun, although sometimes much more active \citep{Peres2000}.
The $R_X$ values exhibit a wide range for stars with similar effective temperatures and surface gravities (also see Figure \ref{lx2lbolT.fig}).
This may represent an intrinsic range of stellar activity,
although it may suffer from some uncertain factors, e.g.,
solar-like cyclic behaviour due to dynamos \citep{Baliunas1995, Olah2002},
long-term variation of stellar activity such as the Maunder minimum phase of solar-type stars
\citep{Baliunas1990, Henry1996, Saar1998, Gray2003},
and variable X-ray emission \citep{Soderblom2010}.
The $R_X$ ranges for each stellar type found in this paper
are consistent with those observed by previous studies \citep{Pizzolato2003, Mamajek2008, Wright2011}.
This suggests that, as expected, the vast majority of these LAMOST spectra are {\it Chandra} source counterparts.

\begin{figure*}[htb]
\center
\includegraphics[width=0.49\textwidth]{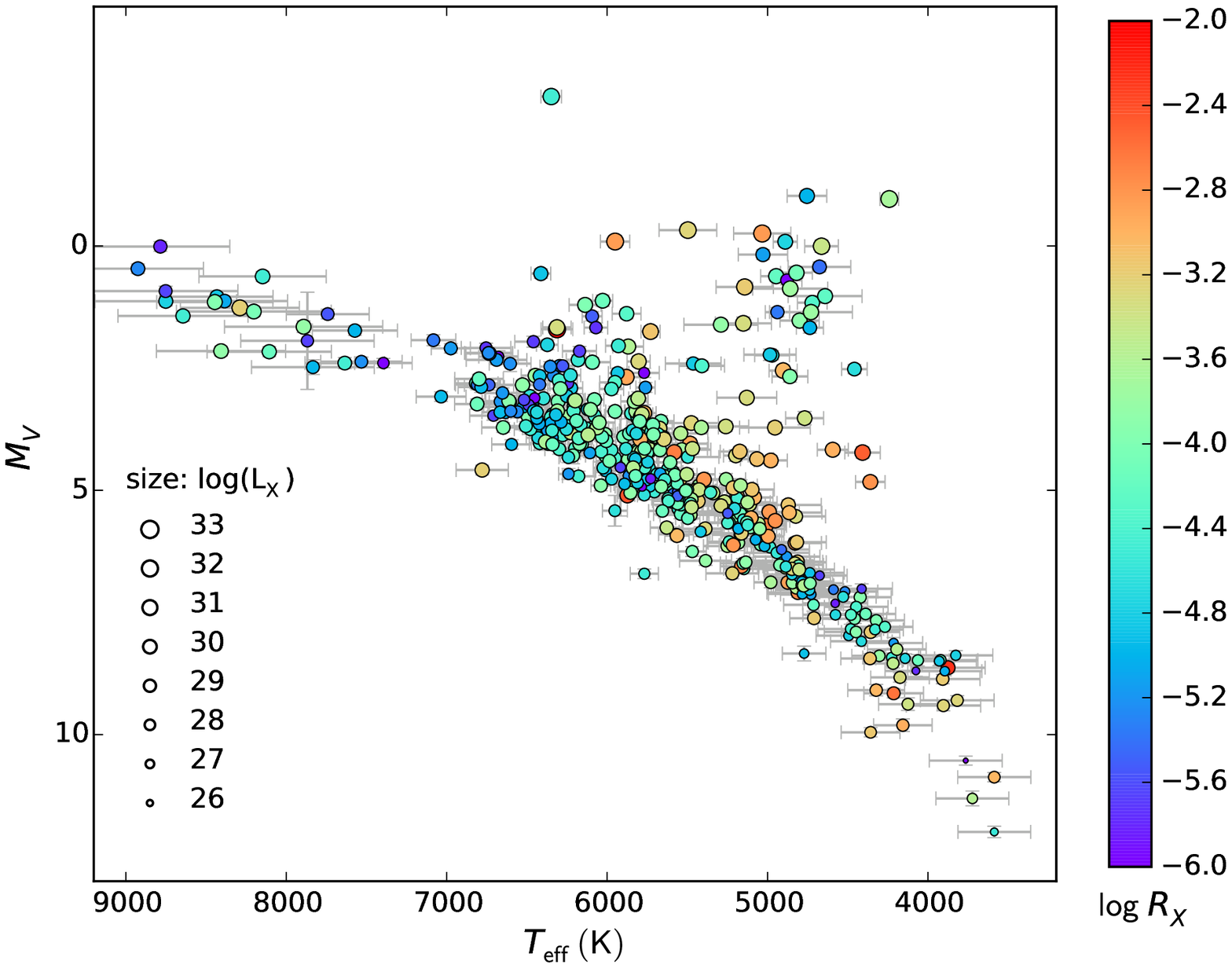}
\includegraphics[width=0.49\textwidth]{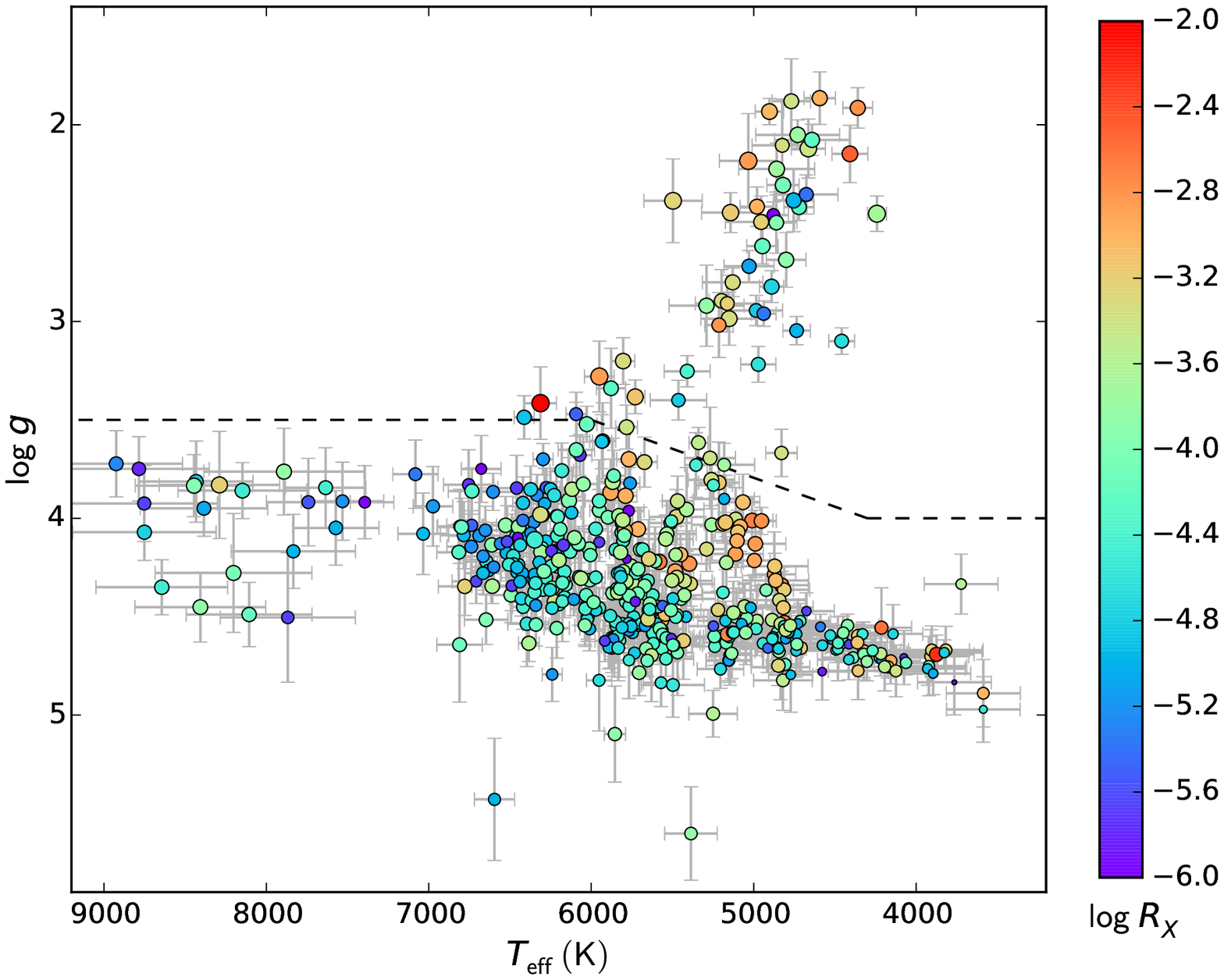}
\caption[]{Left Panel: Hertzsprung-Russell diagram for the sample stars.
The circle size shows the X-ray luminosity, while the color shows the X-ray to optical ratio.
Right Panel: log$g$ versus temperature for the sample sources.
A star is considered to be a giant if it is above the  dashed line,
which is a empirical surface gravity cut separating the giant stars and dwarf stars \citep{Ciardi2011}.}
\label{hr.fig}
\end{figure*}

There are about twenty A-type stars showing X-ray activity.
Generally, A-type stars are thought to be X-ray dark, since
none of the X-ray production mechanisms operating in early- or late-type stars,
i.e., strong winds and convection zones,
ought to be valid for A-type stars.
Previous studies suggested that the X-ray emitting A-types are mostly
pre-main sequence (i.e., Herbig Ae/Be) stars or binaries \citep{Zinnecker1994,Pease2006,Schroder2007}.
In the binary scenario, the X-ray emission is from an unresolved low-mass companion,
often of spectral type dK or dM \citep{Golub1983, Panzera1999}.
We have excluded the identified pre-main sequences and binaries (eclipsing binaries and spectral binaries) from our sample. This does not imply that the X-ray emission from the remaining A-type stars is intrinsic from the stars, because even those sources could still be unrecognized binaries.

From an inspection of the X-ray properties of our sample stars across the Hertzsprung-Russell diagram
(Figure \ref{hr.fig}, left panel),
we find that M dwarfs have the highest $R_X$ but the lowest $L_X$ values;
this confirms the trend previously noted by \citet{Gudel2004}.
There are 51 stars with log$g$ smaller than 3.5, which can be classified as giants.
This confirms the conclusion of previous studies \citep{Simon1989, Auriere2015} that
late-type giants can have (high) stellar activities.
Among these giants with high $R_X$ values, some are very interesting.
For example,
the object J194707.6+445251 is an F5 star with
the highest log$R_X$ value ($\approx-$1.2) among giants.
It has a radio counterpart within a 1$''$.5 radius, called 4C 44.35 \citep{Vollmer2010}.
The possible radio emission suggests the presence of magnetic fields and high-energy electrons \citep{Gudel2002a}.
However, the high X-ray and radio emission are too high for a ``normal" F-type giant.
The X-ray and radio emission are possible to be a chance coincidence with one background AGN, or from the accretion process, which means the F star is in an accreting binary.

\begin{figure}[thp]
\center
\includegraphics[width=0.49\textwidth]{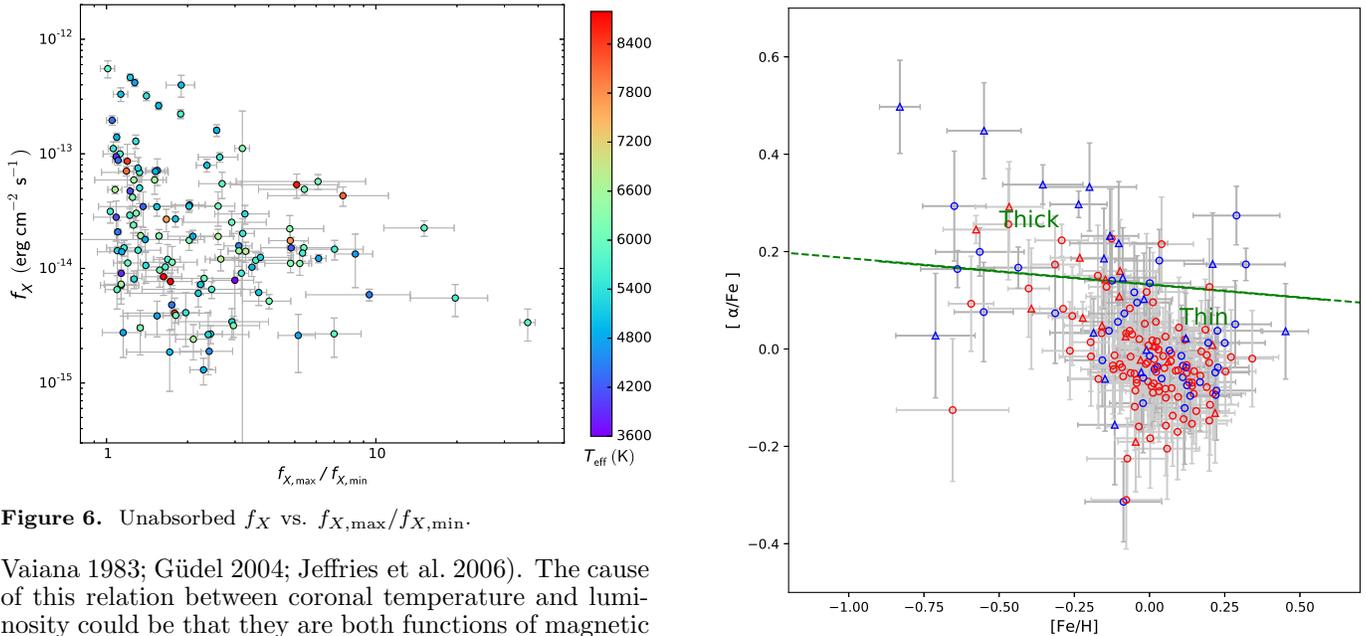}
\caption[]{Unabsorbed $f_X$ vs. $f_{X,\rm max}/f_{X,\rm min}$.}
\label{f2f.fig}
\end{figure}

Stellar X-ray emissions are variable \citep{Soderblom2010},
and X-ray fluxes vary for any spectral type (Figure \ref{f2f.fig}).
In our sample, 126 stars were observed more than once by {\it Chandra}.
About half of them have an X-ray flux variation higher than 2.
The object J131159.5-011705 has the highest flux variation ($\sim$36),
and the variation is not due to a flare event.
The flux variation would affect the correlations between
$R_X$ and other activity indicators and stellar parameters,
which may produce dissimilar results for non-simultaneous observations.

\subsection{The Bimodality of X-ray Activity}
\label{bimodal.sec}

The $R_X$ distributions of G and K stars show clear bimodality (Figure \ref{subclass.fig}).
For G stars, there are more inactive stars than active ones,
while for K stars, more active stars are apparent.
We used double-gaussian functions to fit these $R_X$ distributions:
the peaks for the G type are $-$4.35 and $-$3.3, with $\sigma$ being 0.47 and 0.33;
the peaks for the K type are $-$4.49 and $-$3.18, with $\sigma$ being 0.64 and 0.27.

\begin{figure}[thp]
\center
\includegraphics[width=0.49\textwidth]{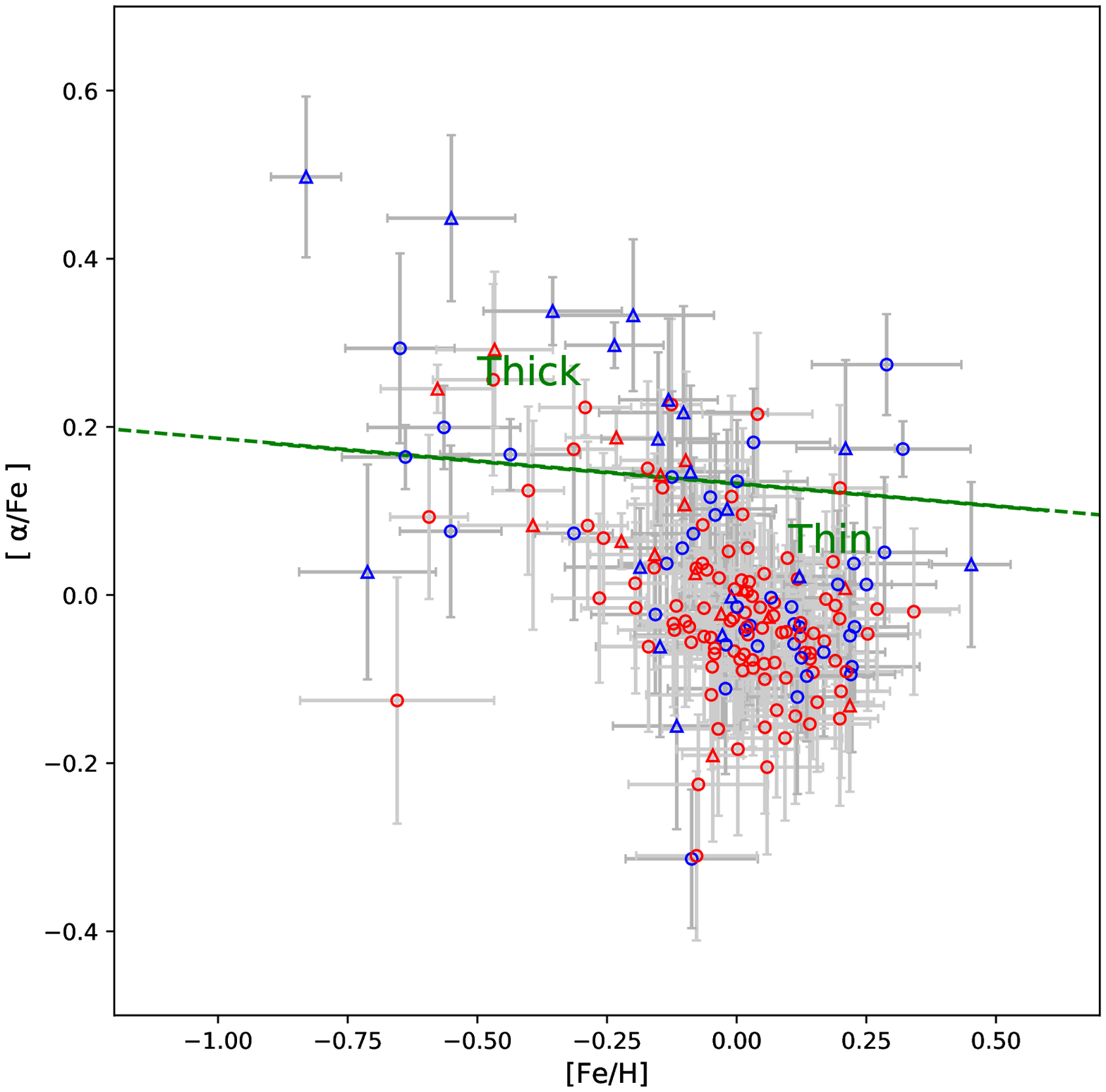}
\caption[]{
[$\alpha$/Fe] versus [Fe/H].
The green line indicates the approximate border of thin disk and thick disk \citep{Masseron2015}.
The circles and triangles represents stars classified as in the thin disk and thick disk, respectively,
using the kinematic methods.}
\label{feh.fig}
\end{figure}

The {\it Chandra} data are drawn from targeted observations, which means that our sample is inhomogeneous.
Therefore, it is necessary to check whether the bimodality is
a result of the sampling effect,
e.g., two populations of stars in different parts of the Galaxy.
We divided the G- and K-type stars into two sub-samples, corresponding to stars with log$R_X > -3.8$ and log$R_X < -3.8$.
To identify whether one source belongs to the thin disk or thick disk,
we first defined one probability assuming that the velocities
($U_{LSR}$, $V_{LSR}$, $W_{LSR}$)
follow a 3-D Gaussian distribution \citep{Guo2016}:
\begin{equation}
{\rm Prob} = c{\cdot}{\rm exp}\{-\frac{U_{\rm LSR}^2}{2\sigma_{U}^2} -\frac{(V_{\rm LSR}- V_{\rm asym})^2}{2\sigma_{V}^2} -\frac{W_{\rm LSR}^2}{2\sigma_{W}^2} \},
\end{equation}
where $c = (2\pi)^{-3/2}(\sigma_{U}\sigma_{V}\sigma_{W})^{-1}$ normalizes the expression.
$V_{\rm asym}$ is the asymmetric drift, and $\sigma_{U}$, $\sigma_{V}$, and $\sigma_{W}$
are the velocity dispersions in three dimensions, all of which
varies with different components \citep{Guo2016}.
Then we defined the probability ratio of belonging to the thin disk or thick disk as
\begin{equation}
f_{\rm thin/thick} = \frac{\rm Prob_{thin/thick}}{\rm Prob_{thin} + Prob_{thick}+Prob_{halo}}.
\end{equation}
We classified a star as being a good candidate star in the thin disk or thick disk when this ratio is larger than 80\%.
Figure \ref{feh.fig} shows the classification of the sample stars in the [$\alpha$/Fe]-[Fe/H] diagram.
The circles and triangles represents stars classified as in the thin disk or thick disk,
using the kinematic methods.
The green dashed line indicates the approximate border of thin disk and thick disk from metallicity \citep{Masseron2015}.
Most stars, both active and inactive, are identified as being in the thin disk.
Therefore, both the metallicity and velocity information show that
the bimodality is not caused by different location of the stars, i.e. the sampling effect.

Previous studies with Ca II HK,  H$\alpha$, and X-ray emission
have found the bimodality of stellar activity, with an active and inactive peak
\citep[e.g.,][]{Stocke1991, Henry1996, Wright2004, Jenkins2006, Jenkins2008, Jenkins2011,Agueros2009, Martinez-Arnaiz2011,Pace2013}.
The bimodality is explained as one saturated and one non-saturated subpopulation \citep{Martinez-Arnaiz2011,Katsova2016},
and can further be explained as one young and one old subpopulation.
The latter one is often thought to be inactive in chromospheric and X-ray emission, since
the magnetic activity decreases simultaneously as the rotation decelerates with age
\citep[e.g.,][]{Mamajek2008,Katsova2011}.

\begin{figure}[thp]
\center
\includegraphics[width=0.49\textwidth]{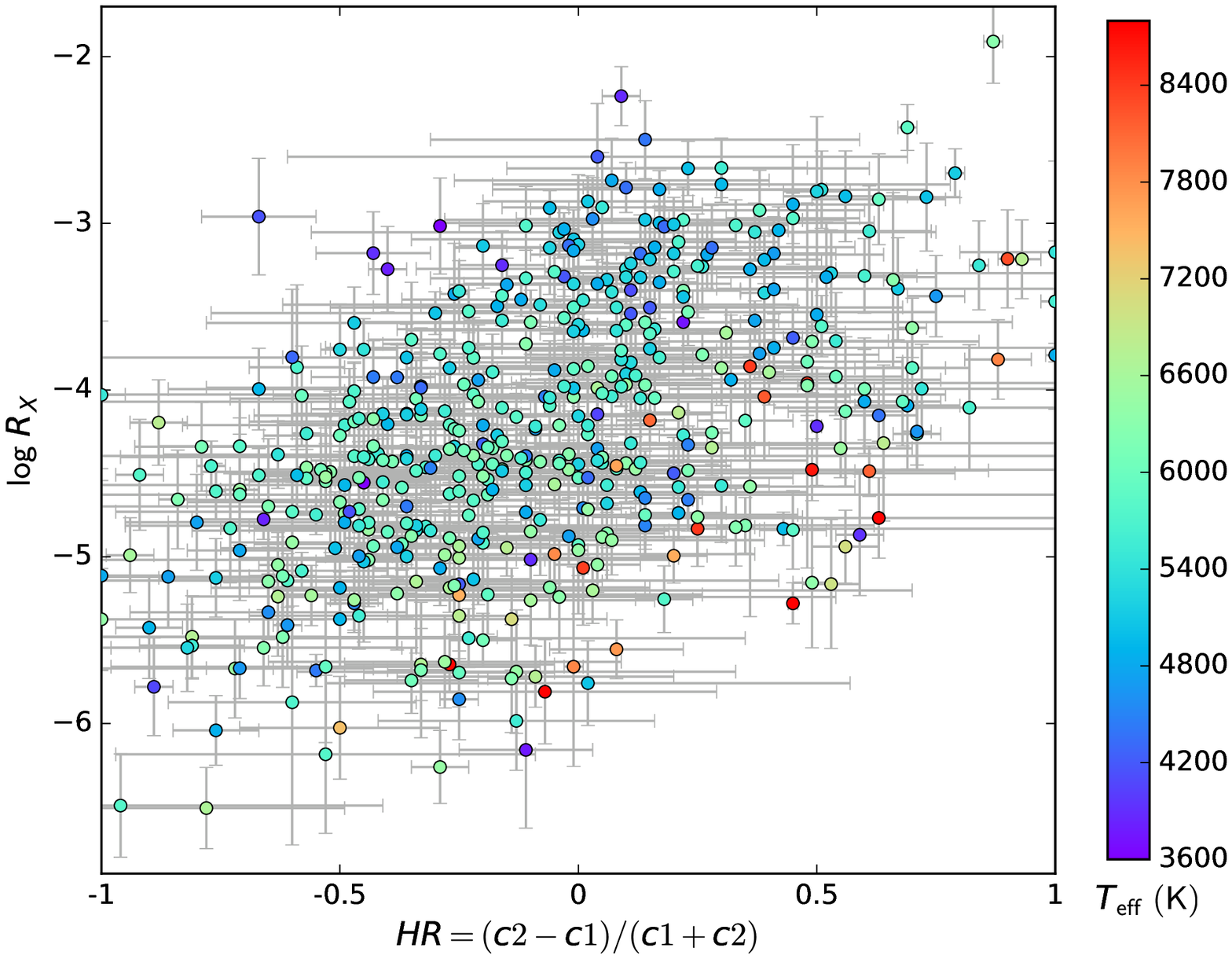}
\caption[]{$R_X$ vs. hardness ratio $HR$.
The positive correlation means stronger X-ray emitters (higher $R_X$) have higher coronal temperatures.}
\label{color1.fig}
\end{figure}

\begin{figure}[htb]
\center
\includegraphics[width=0.49\textwidth]{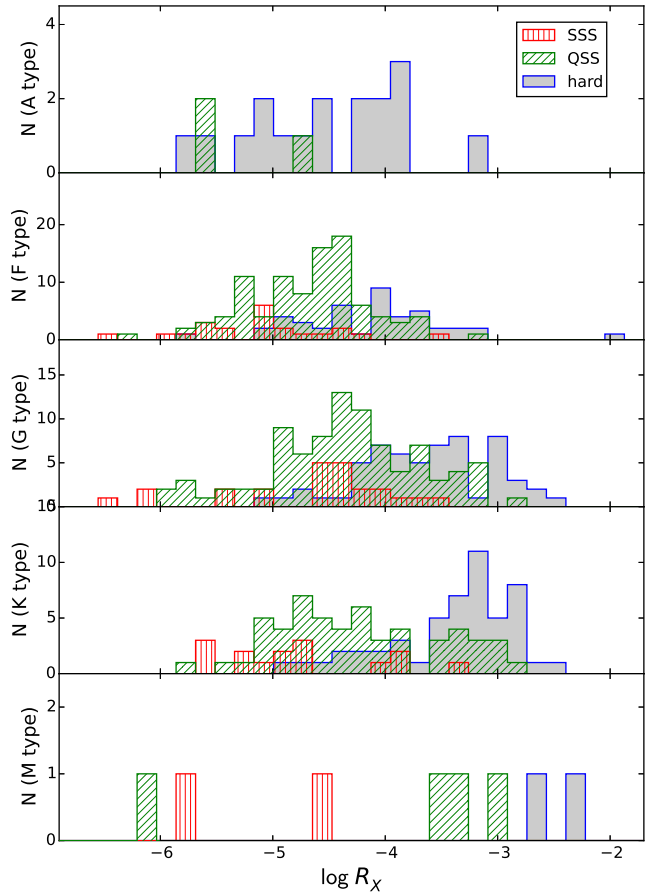}
\caption[]{Histograms showing the distributions of $R_X$ for stars from A to M type.
These histograms plotted with red vertical lines, green slashes, and blue shaded area indicate
SSSs, QSSs, and hard sources, respectively.
Generally sources with harder X-ray spectra have higher $R_X$ values.}
\label{color2.fig}
\end{figure}

The coronal temperature is known to be positively correlated with X-ray luminosity and stellar activity
\citep[e.g.,][]{Vaiana1983, Gudel2004, Jeffries2006}.
The cause of this relation between coronal temperature and luminosity could be that they are both functions of magnetic activity \citep{Gudel2004}. A more efficient dynamo inside active stars (for example because of faster rotation) produces stronger magnetic fields in the corona, and consequently a higher rate of field line reconnections and flares. This results both in a larger density of energetic electrons in the corona, and in higher temperatures.
Therefore, we suggest that the double-peaked distribution of $R_X$
represent a double-peaked distribution of heating rates, and therefore coronal temperatures.
We do not have direct measurements of coronal temperatures for the stars in our sample;
however, we take the X-ray hardness ratio as a proxy for the coronal temperature,
because the thermal plasma emission shifts progressively to higher photon energies for higher plasma temperatures.
There is a clear positive correlation between $R_X$ and $HR$ (Figure \ref{color1.fig}), whcih means stronger X-ray emitters (higher $R_X$) have higher coronal temperatures.
{\it Chandra} sources are sometimes empirically classified into three groups: super-soft, quasi-soft, and hard, based on their hardness ratios
\citep{Di Stefano2003}. Applying the same classification to our sample stars, we find (Figure \ref{color2.fig}) that the higher activity peak
of the $R_X$ distribution in G- and K-type stars is dominated by hard sources,
and the lower activity peak by quasi-soft and super-soft sources.

\begin{figure*}[htb]
\center
\includegraphics[width=0.49\textwidth]{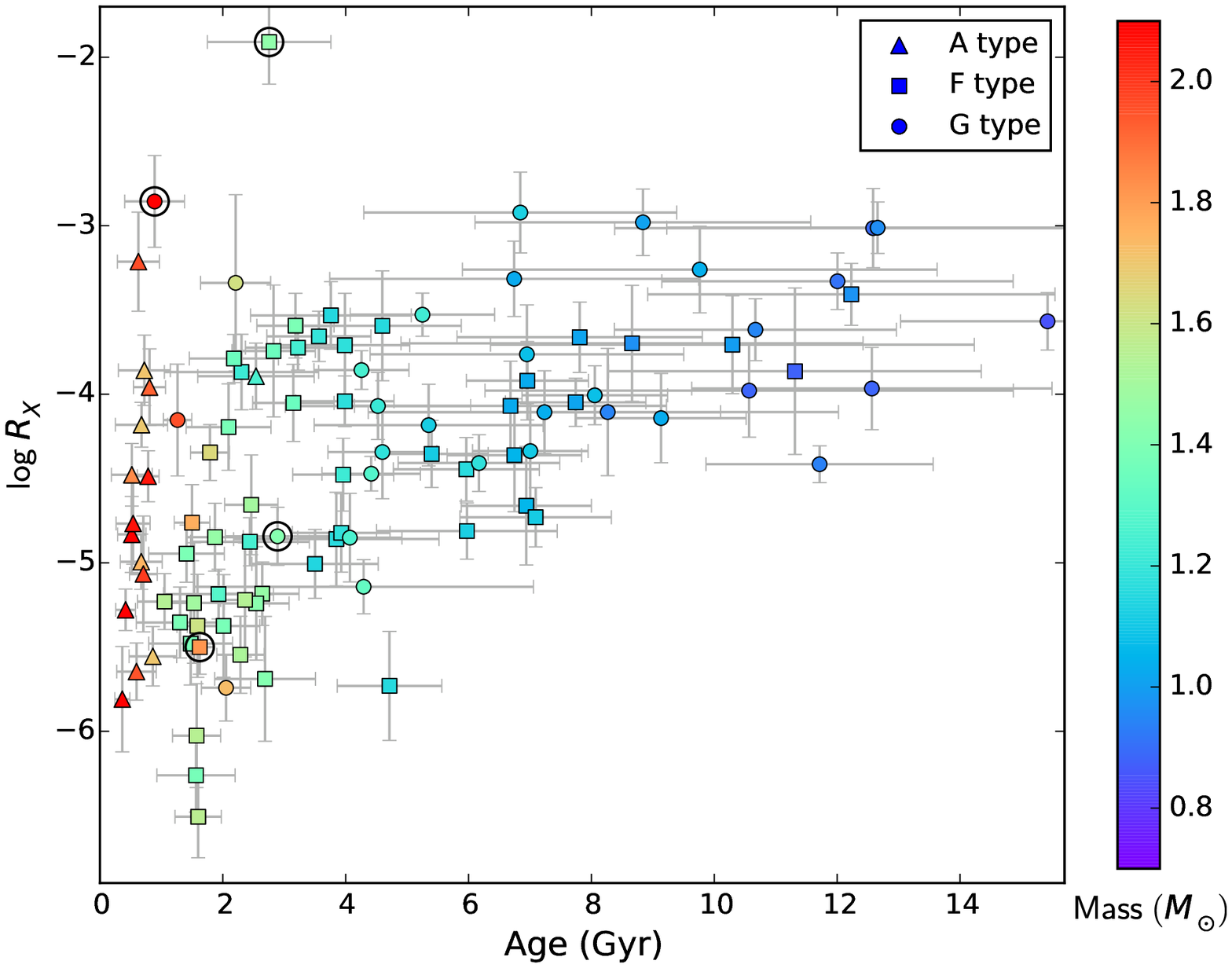}
\includegraphics[width=0.49\textwidth]{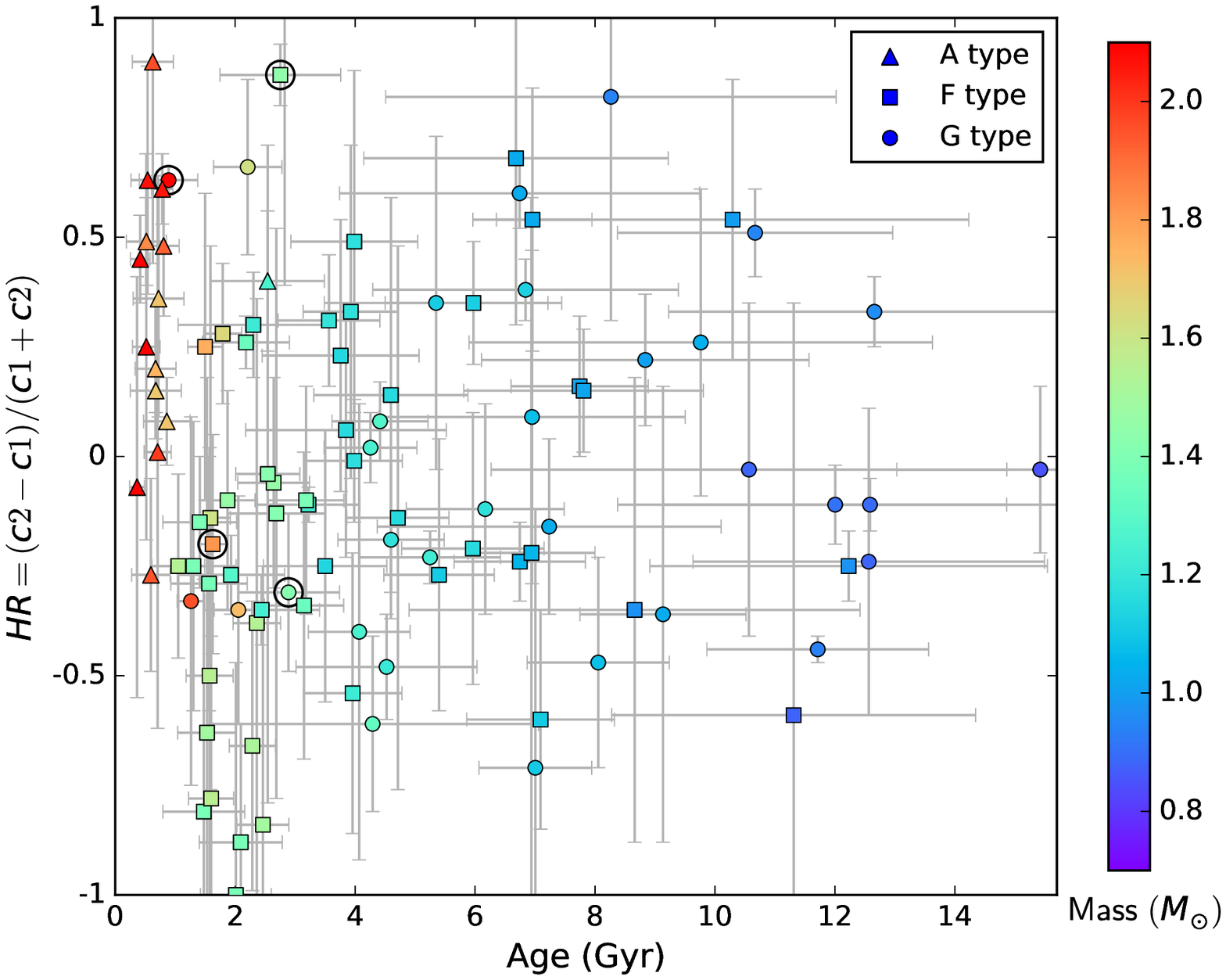}
\caption[]{Left Panel: $R_X$ vs. stellar age.
A trough with lowest $R_X$ values around 2 Gyr seems to be consistent with \citet{Pace2013}.
However, no trend can be confirmed due to the uncertainty of the ages.
The A, F, and G types are plotted with triangles, squares, and circles, respectively.
The four objects plotted with larger black circles are giants.
Right Panel: $HR$ as a function of stellar age.}
\label{agemass.fig}
\end{figure*}

\subsection{The Evolution of X-ray Activity}
\label{age.sec}

Previous studies have shown that
X-ray luminous, single, low-mass stars are the youngest \citep[e.g.,][]{Telleschi2007},
while older stars have typically lower X-ray luminosities \citep{Flaccomio2003, Feigelson2004, Wright2010}.
This means that the X-ray emission decreases over the lifetime of the star.
This is explained as a consequence of the stellar dynamo
\citep[e.g.,][]{Skumanich1972, Flaccomio2003, Feigelson2004, Cranmer2011}:
over time, a magnetized stellar wind driven by the stellar dynamo
leads to the loss of angular momentum and the rotational spin-down of the star,
which reduces velocity shear at the tachocline between the radiative and convective zones,
resulting in a reduced magnetic field (and therefore reduced activity).

Recently, many new results have been obtained about the age-activity relation.
An $L$-shaped relation was found between the chromospheric activity ($R^{'}_{HK}$) and age  \citep{Pace2013}:
the chromospheric activity decreases with age for stars younger than 2 Gyr,
and there is no decay of chromospheric activity after about 1.5--2 Gyr \citep{Lyra2005, Sissa2016}.
However, using SDSS observations of open clusters,
\citet{Zhao2013} found that the chromospheric activity gradually declines with age, extending to at least $\sim$8 Gyr.
A steepening of the relation between activity and rotation was reported for stars with ages between 1 and 10 Gyr,
but was possibly attributed to a non-representative sample of a few target objects.
Several improved age-activity relations extending to old late-type stars have also been obtained
\citep{Mamajek2008,Zerjal2017}.

\citet{Xiang2017a} estimated ages and masses for 0.93 million main-sequence turnoff and subgiant stars,
by matching with stellar isochrones employing a Bayesian algorithm, using effective temperatures, absolute magnitudes,
metallicities ([Fe/H]), and $\alpha$-element to iron abundance ratios ([$\alpha$/Fe]) deduced from the LAMOST spectra.
There are 16 A types, 60 F types, and 34 G types with estimated ages and masses in our sample.
We do not find a clear trend between $R_X$ and age (Figure \ref{agemass.fig}).
A trough with the lowest $R_X$ values around 2 Gyr is consistent with \citet{Pace2013},
but the reason is not clear.
There are so few sources younger than 1 Gyr that we cannot confirm whether
the activity keep decreasing over the first several hundred Myrs.
There is also no clear trend between $HR$ and age.

\begin{table}
\begin{center}
\scriptsize
\setlength{\tabcolsep}{2pt}
\caption[]{Stars with ${\rm H\alpha}$ emission lines in our sample.}
\label{emission.tab}
\begin{tabular}{lcccc}
\hline\noalign{\smallskip}
Object  & EW$_{\rm H\alpha}$  & EW$'_{\rm H\alpha}$  & log$R_{\rm H\alpha}$  \\
        &  $\AA$     &     $\AA$    &               \\
  (1)   &     (2)    &    (3)       &   (4)          \\
\hline\noalign{\tiny}
J004322.0+405752 & 1.95 $\pm$ 0.19 & 3.43 & $-$3.5 $\pm$ 0.05\\
J004458.9+330431 & 0.51 $\pm$ 0.13 & 1.93 & $-$3.72 $\pm$ 0.08\\
J013729.6+331230 & 1.82 $\pm$ 0.1 & 2.63 & $-$3.64 $\pm$ 0.1\\
J022641.9+002811 & 0.35 $\pm$ 0.02 & 2.2 & $-$3.59 $\pm$ 0.04\\
J024132.9+002053 & 2.48 $\pm$ 0.05 & 3.42 & $-$3.48 $\pm$ 0.08\\
J033102.5+434758 & 0.65 $\pm$ 0.05 & 2.41 & $-$3.62 $\pm$ 0.06\\
J034501.5+321050 & 0.78 $\pm$ 0.06 & 2.39 & $-$3.58 $\pm$ 0.07\\
J034631.1+240702 & 1.64 $\pm$ 0.05 & 2.45 & $-$3.69 $\pm$ 0.1\\
J040355.6+261652 & 1.35 $\pm$ 0.06 & 2.13 & $-$3.93 $\pm$ 0.11\\
J054143.2$-$020353 & 1.03 $\pm$ 0.07 & 2.46 & $-$3.62 $\pm$ 0.06\\
J054211.2$-$015943 & 0.85 $\pm$ 0.05 & 2.33 & $-$3.66 $\pm$ 0.07\\
J063005.3+054540 & 2.1 $\pm$ 0.07 & 3.19 & $-$3.46 $\pm$ 0.04\\
J063324.2+222837 & 8.96 $\pm$ 0.33 & 9.64 & $-$3.29 $\pm$ 0.11\\
J071618.9+374451 & 1.69 $\pm$ 0.12 & 3.07 & $-$3.49 $\pm$ 0.05\\
J095817.1+362216 & 1.17 $\pm$ 0.1 & 1.9 & $-$3.89 $\pm$ 0.11\\
J095948.8$-$051413 & 0.93 $\pm$ 0.07 & 2.26 & $-$3.63 $\pm$ 0.06\\
J111032.9+570633 & 1.99 $\pm$ 0.02 & 3.5 & $-$3.39 $\pm$ 0.03\\
J111637.2+013249 & 1.26 $\pm$ 0.05 & 2.29 & $-$3.72 $\pm$ 0.07\\
J122837.2+015720 & 2.19 $\pm$ 0.03 & 3.16 & $-$3.54 $\pm$ 0.08\\
J131156.7$-$011311 & 0.78 $\pm$ 0.03 & 2.25 & $-$3.66 $\pm$ 0.07\\
J203323.0+411222 & 1.69 $\pm$ 0.05 & 3.25 & $-$3.44 $\pm$ 0.03\\
J203552.9+411503 & 1.39 $\pm$ 0.05 & 2.31 & $-$3.79 $\pm$ 0.08\\
J223606.9+012603 & 1.31 $\pm$ 0.05 & 2.53 & $-$3.65 $\pm$ 0.04\\
J224506.0+394016 & 0.98 $\pm$ 0.05 & 2.4 & $-$3.62 $\pm$ 0.07\\
\noalign{\smallskip}\hline
\end{tabular}
\end{center}
\end{table}

\begin{figure*}[htb]
  \center
\includegraphics[width=0.49\textwidth]{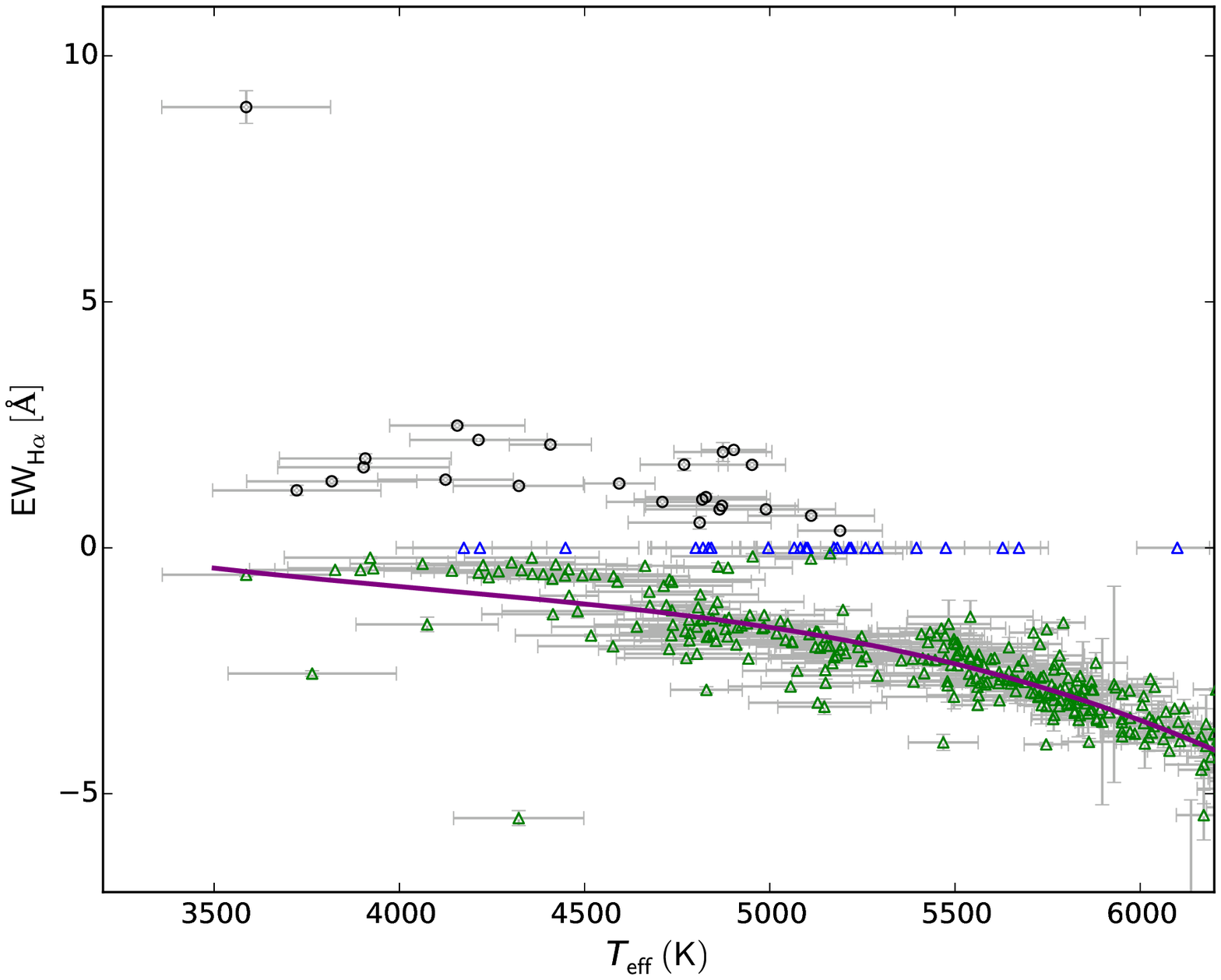}
\includegraphics[width=0.48\textwidth]{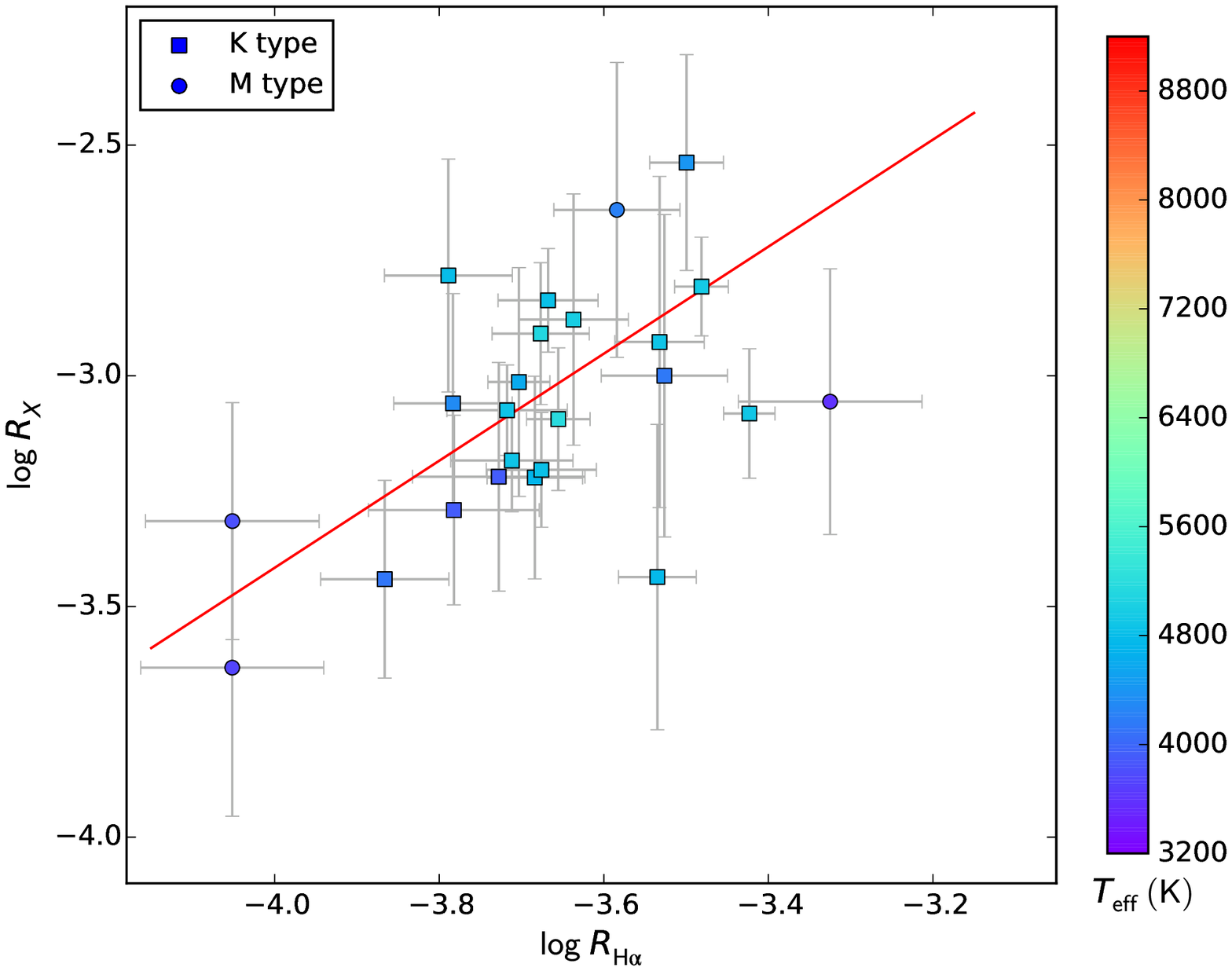}
\caption[]{Left Panel:
EW$_{\rm H\alpha}$ as a function of $T_{\rm eff}$.
The black circles, blue triangles, green triangles represent EW values as positive, zero, and negative, respectively.
The solid line is the fitted ``basal line'' using the stars (3500 K $< T_{\rm eff} < $ 6200 K) with negative EWs.
Right Panel:
  Relation between $R_X$ and normalized H$\alpha$ luminosity.
The K and M types are plotted with squares and circles, respectively.
The red line is a linear fitting to the data as
${\rm log}~R_X = (1.16 \pm 0.28)\times~{\rm log}~R_{\rm H\alpha} + (1.22 \pm 1.03)$.}
\label{hahb.fig}
\end{figure*}

\subsection{Relation Between X-ray and Chromospheric Activity}
\label{ew.sec}

The X-ray emission and other activity indicators are proxies of stellar magnetic activity \citep{Testa2015}.
The X-ray activity shows a good correlation with the activity indicators of chromosphere and transition region,
e.g., Ca II HK emission \citep{Vaiana1983,Schrijver1992,Sterzik1997},
Mg II emission of dMe stars \citep{Mathioudakis1989},
and H$\alpha$ and UV emission of M dwarfs \citep{Martinez-Arnaiz2011,Stelzer2012, Stelzer2013}.
LAMOST provides us with a good opportunity to study the relation between X-ray activity and H$\alpha$ emission.

First, we calculated the equivalent widths (EWs) of the H$\alpha$ emission line.
The EW was calculated using the following formula:
\begin{equation}
{\rm EW} = \int \frac{f(\lambda)-f(0)}{f(0)} d\lambda,
\label{ew.eq}
\end{equation}
where $f(0)$ denotes the nearby pseudo-continuum flux. 
In order to trace pure chromospheric emission,
we calculated the excess EW (hereafter EW$'$) by subtracting a ``basal line'' of chromospheric emissions \citep{Fang2018},
which is constructed using those stars with H$\alpha$ absorption lines (EW $<$ 0).
Then, we
used the CK04 stellar atmosphere models \citep{ck04} to transform
the EW$'$s to stellar surface fluxes of H$\alpha$ emission lines, $f_{\rm H\alpha, sur}$ \citep{Yang2017}.
For each star, the model with the most similar $T_{\rm eff}$, log$g$, and [Fe/H] was used.
Finally, we determined the flux ratio $f_{\rm H\alpha, sur}/f_{\rm bol, sur}$ using $f_{\rm bol, sur}=\sigma \,{T}^{4}$,
and calculated the H$\alpha$ emission index, $R_{\rm H\alpha} = L_{\rm H\alpha}/L_{\rm bol} = f_{\rm H\alpha, sur}/f_{\rm bol, sur}$,
to quantify stellar chromospheric activity (Table \ref{emission.tab}).

Figure \ref{hahb.fig} shows a general power law dependence of $R_X$ on $R_{\rm H\alpha}$:
\begin{equation}
{\rm log}~R_X = (1.16 \pm 0.28)\times~{\rm log}~R_{\rm H\alpha} + (1.22 \pm 1.03).
\end{equation}
There are 20 K-type and 4 M-type stars in the $R_X$--$R_{\rm H\alpha}$ diagram.
We propose the scatter in the relations is caused by the non-simultaneous X-ray and H$\alpha$ observations.
However, \citet{Stelzer2013} reported that the scatter may represent an intrinsic range
of physical conditions between the corona and chromosphere.
It was reported by \citet{Martinez-Arnaiz2011} that in a sample of late-type dwarf active stars with spectral types from F to M, $F_X \propto F_{\rm H\alpha}^{1.48\pm0.07}$. For M dwarfs, \citet{Stelzer2013} derived $R_X \propto R_{\rm H\alpha}^{1.90\pm0.31}$. In this paper, we have obtained that $R_X \propto R_{\rm H\alpha}^{1.12 \pm 0.30}$, a slightly flatter relation than found in those other studies. The discrepancy may be due to our small sample size. In addition, we are aware that the lack of simultaneous observations of those two intrinsically varying properties (coronal and chromospheric activities) introduces another source of uncertainty in all these studies \citep{Martinez-Arnaiz2011}.

\subsection{Stars with X-ray Flares}
\label{superflare.sec}

Flares are in the center of the debate on the origin of coronal heating \citep{Audard2000},
with evidence showing that (X-ray or $U$-band) flares can release a sufficient amount of energy
to produce quiescent coronal emission
\citep{Doyle1985,Skumanich1985,Pallavicini1990,Pandey2008,Pye2015}.
The light curves of flares can help us understand the characteristics of the coronal structures and,
therefore, of the magnetic field \citep{Schmitt1999,Favata2000,Reale2004,Pandey2008}

The X-ray emission from flares is indicative of very energetic, transient phenomena,
associated with energy release via magnetic reconnection \citep{Pye2015}.
X-ray flares differ considerably among different stars \citep{Benz2010},
e.g., in shape \citep{Cully1993,Graffagnino1995,Osten1999},
peak luminosity ($\approx$10$^{26}$--10$^{33}$ erg s$^{-1}$)
and total energy ($\approx$10$^{28}$--10$^{37}$ erg)
\citep{Kuerster1996,Gudel2002,Osten2007,Getman2008},
and rise and decay times \citep{Haisch1987, Tagliaferri1991, Cully1994, Benz2010}.

\begin{figure*}[htb]
\includegraphics[width=0.33\textwidth]{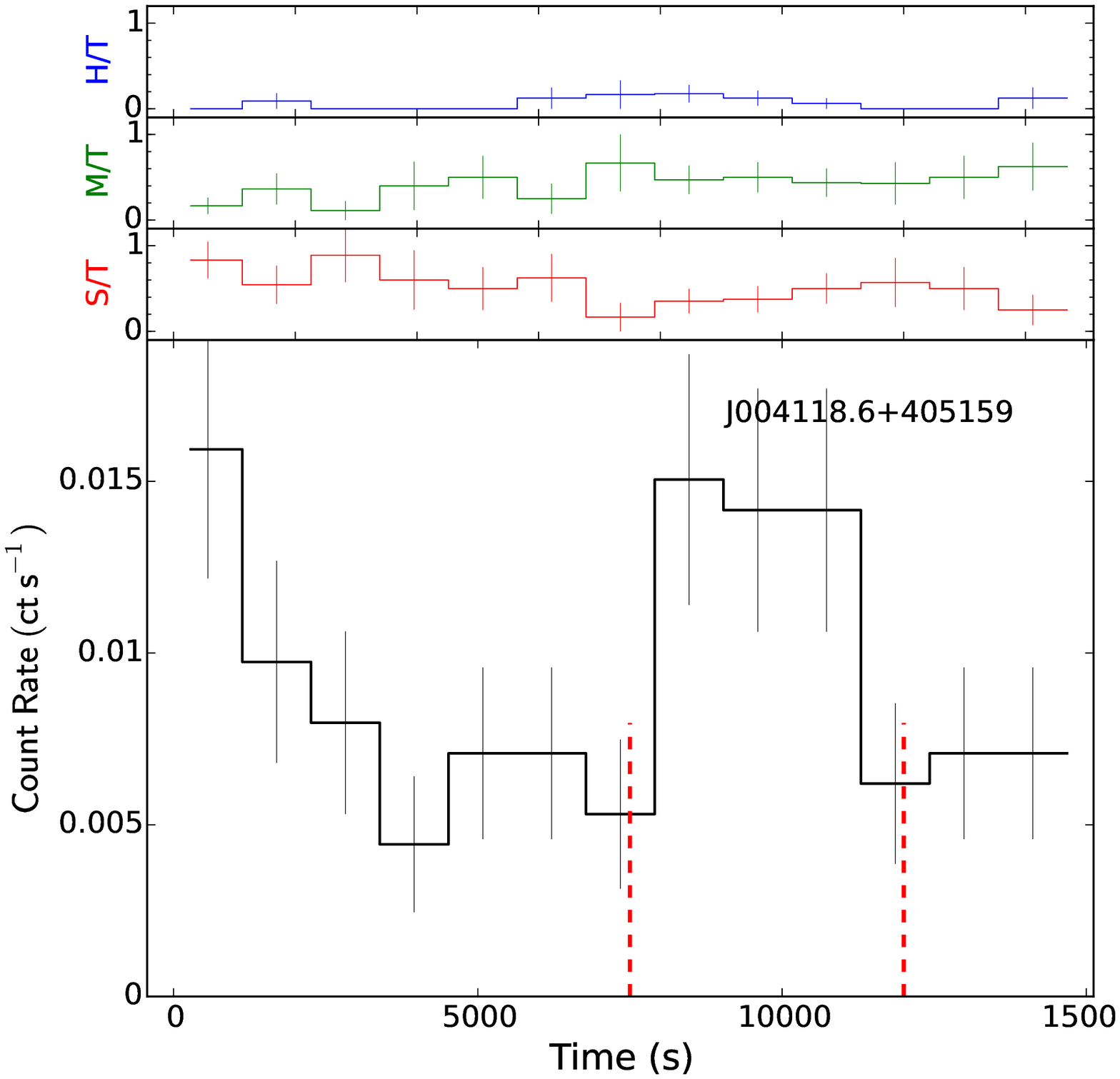}
\includegraphics[width=0.33\textwidth]{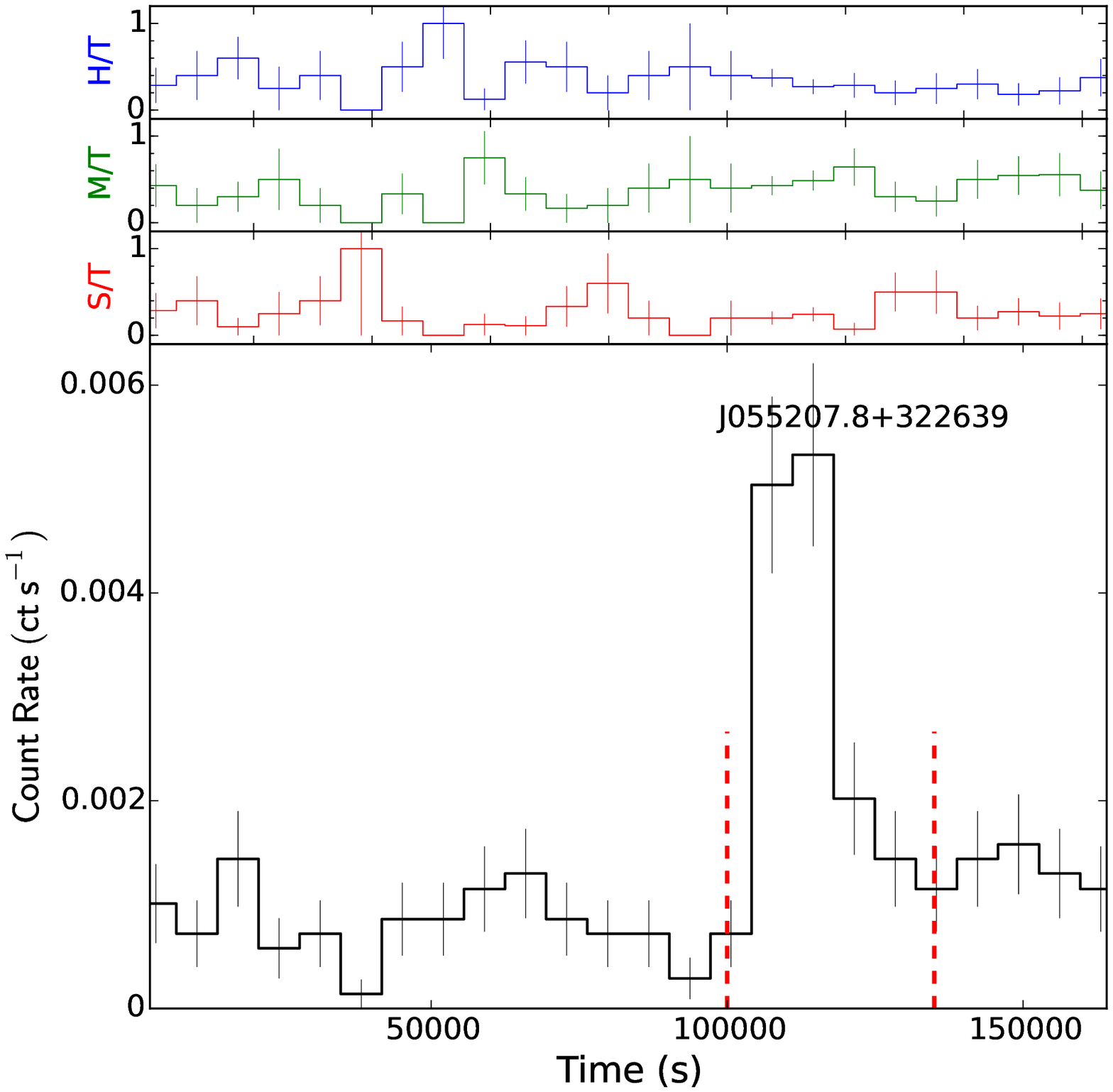}
\includegraphics[width=0.33\textwidth]{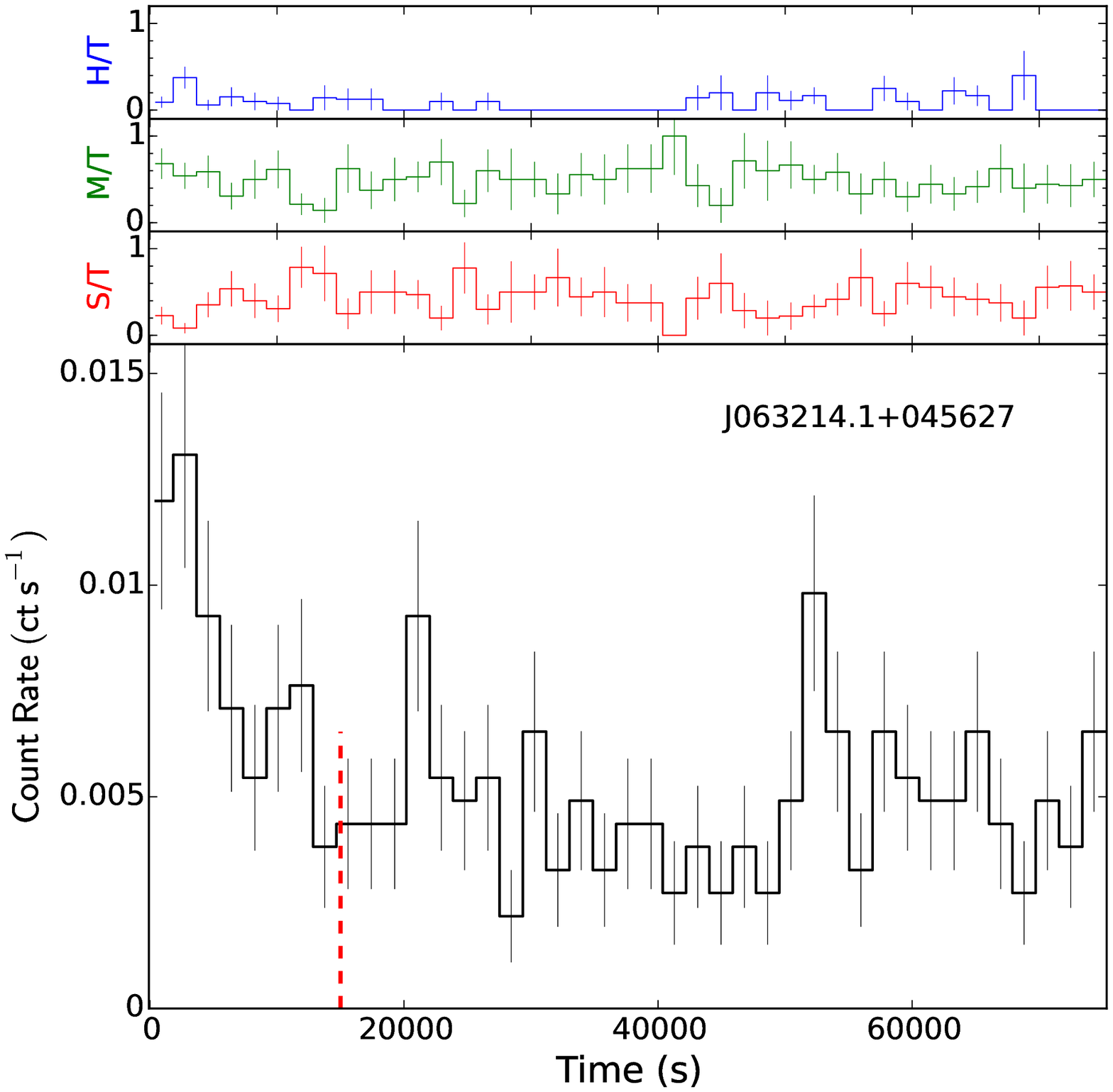} \\
\includegraphics[width=0.33\textwidth]{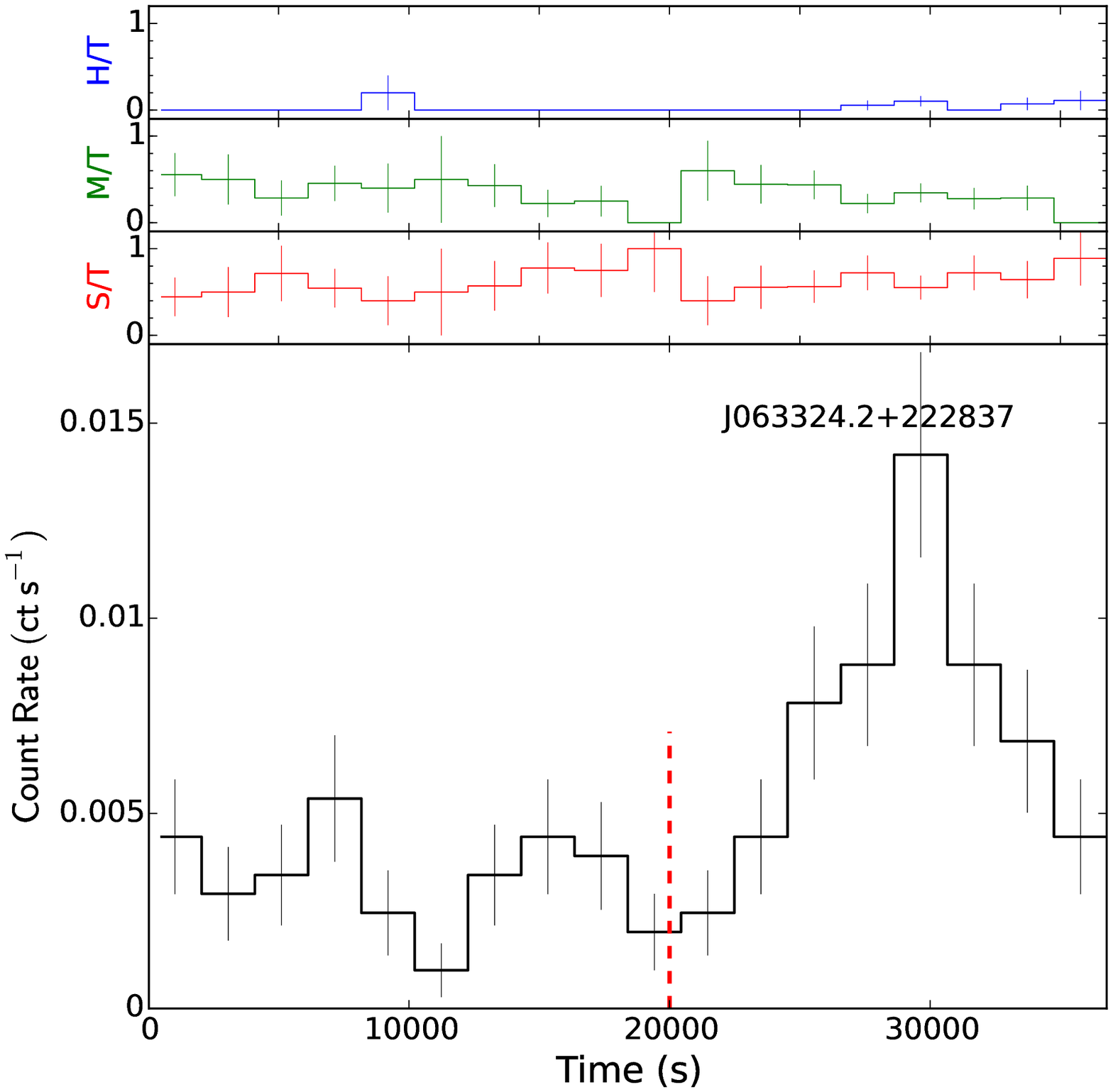}
\includegraphics[width=0.33\textwidth]{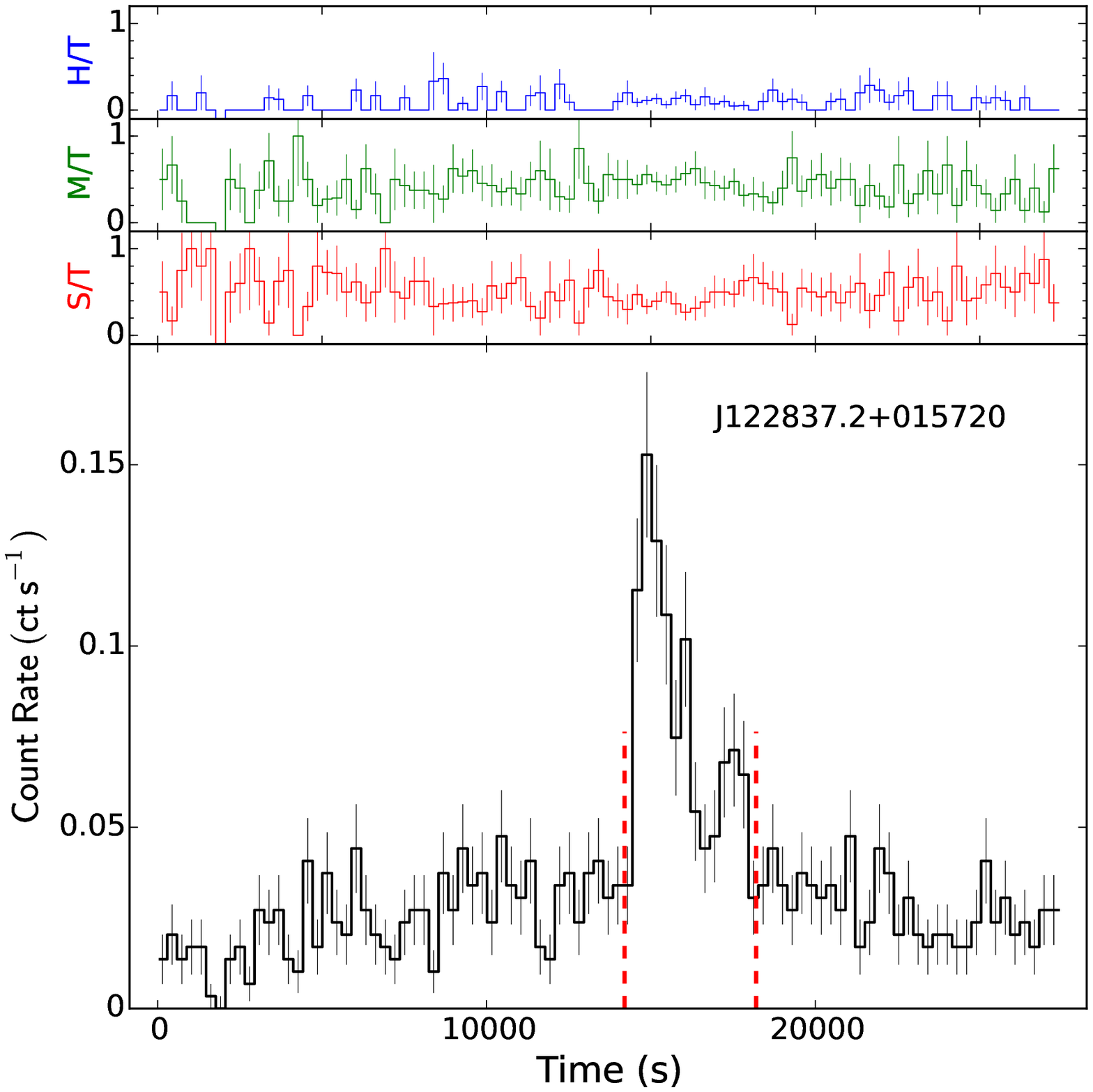}
\includegraphics[width=0.33\textwidth]{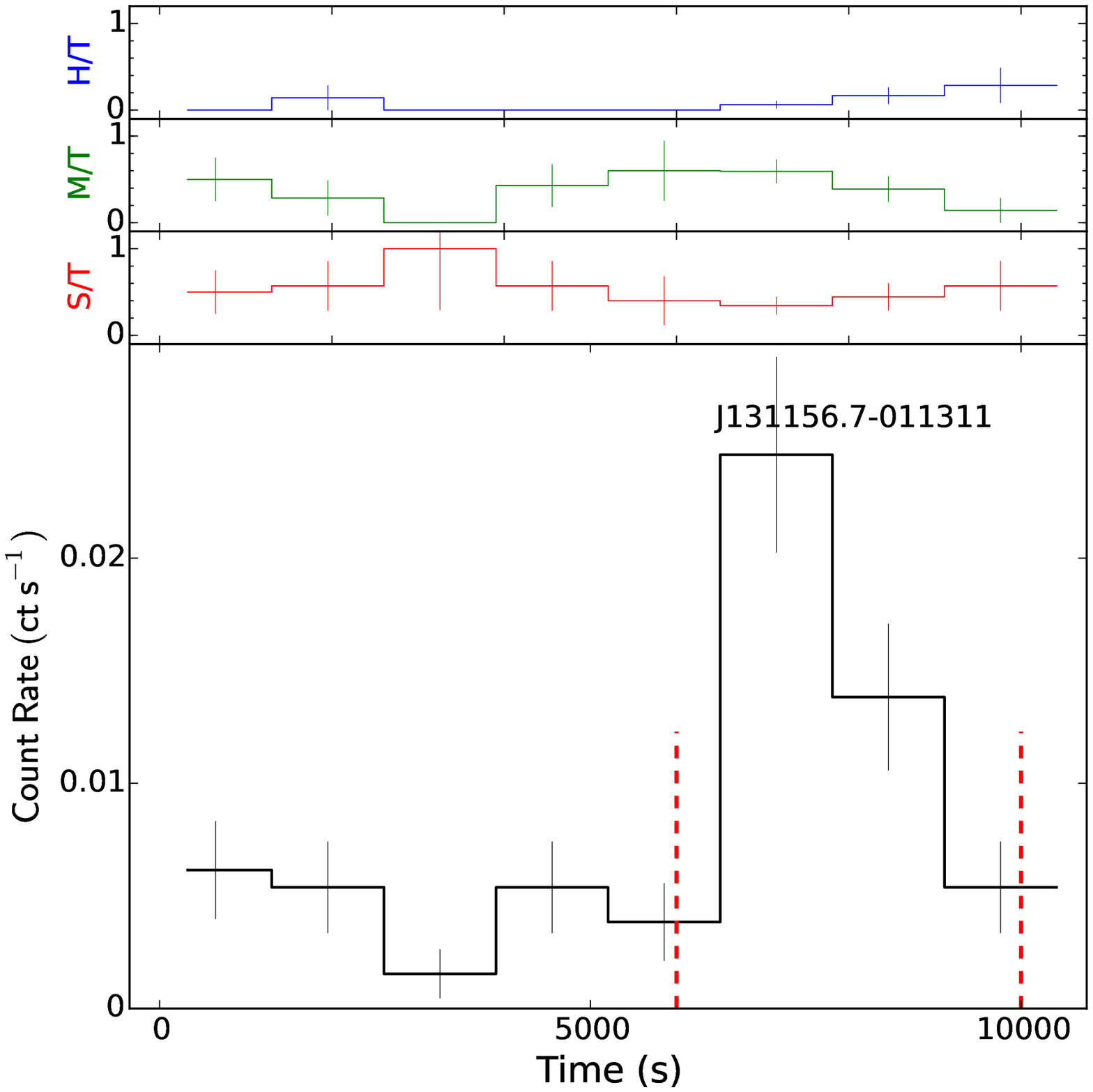}\\
\includegraphics[width=0.33\textwidth]{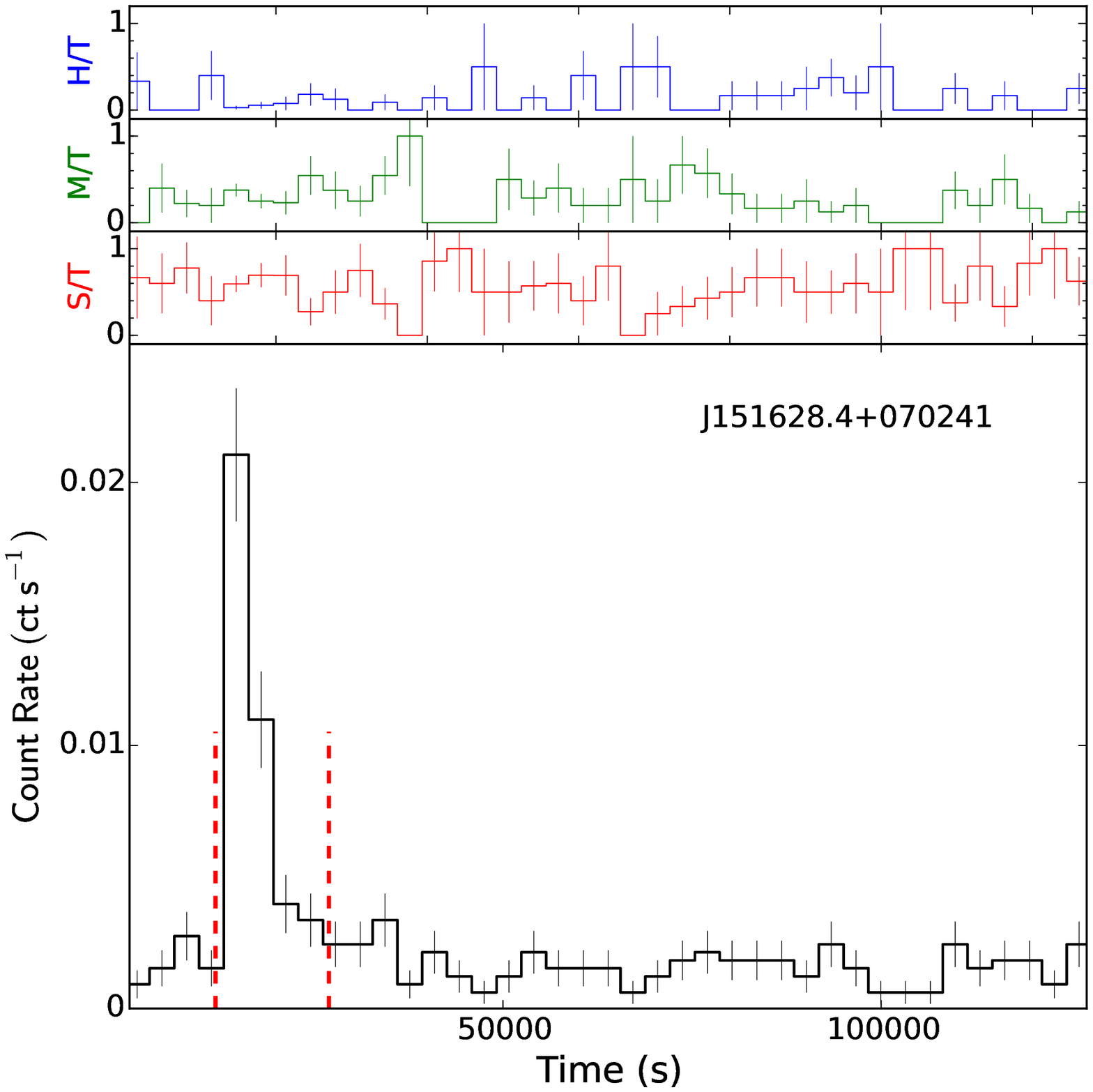}
\includegraphics[width=0.33\textwidth]{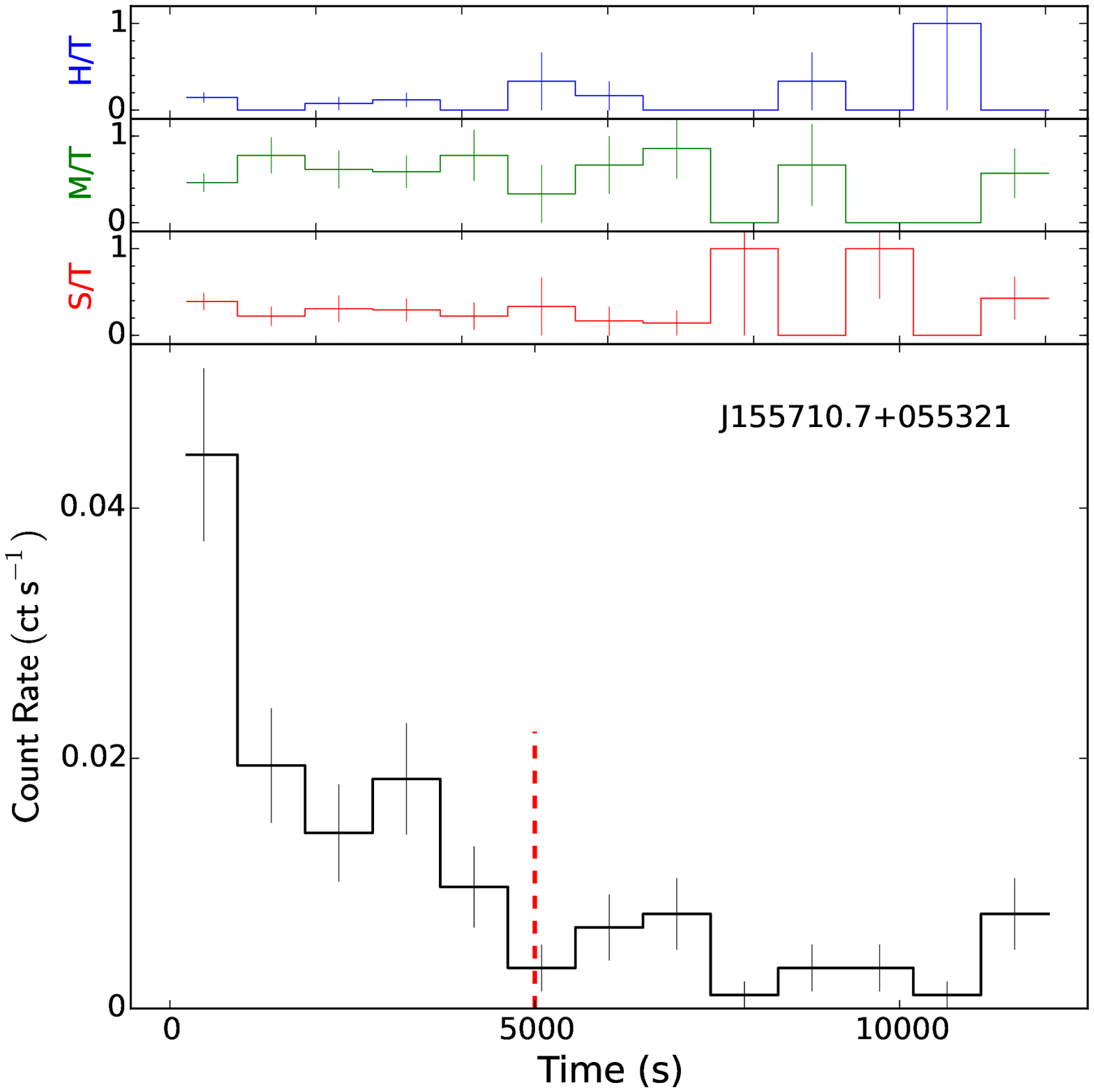}
\includegraphics[width=0.33\textwidth]{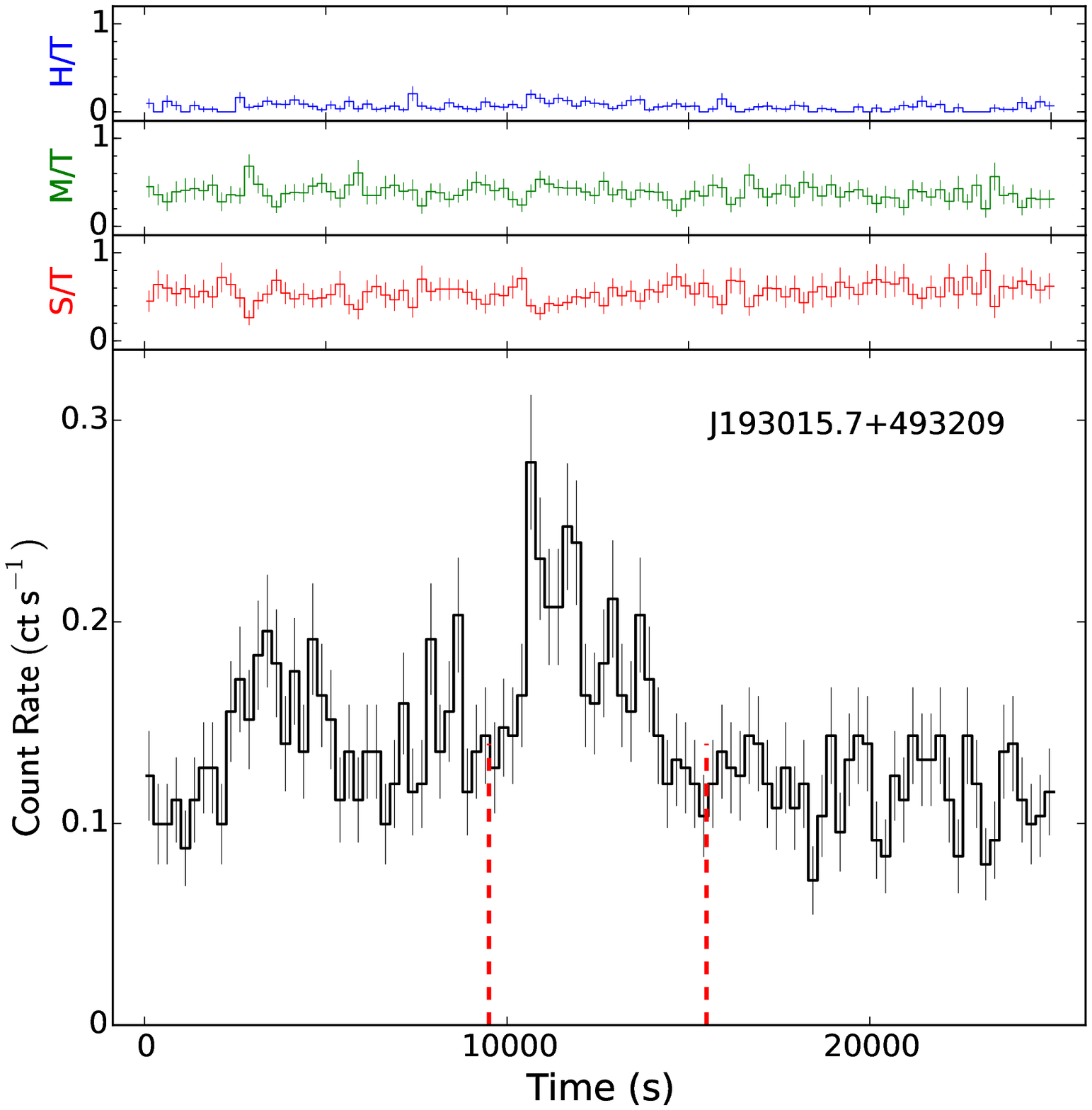}\\
\includegraphics[width=0.33\textwidth]{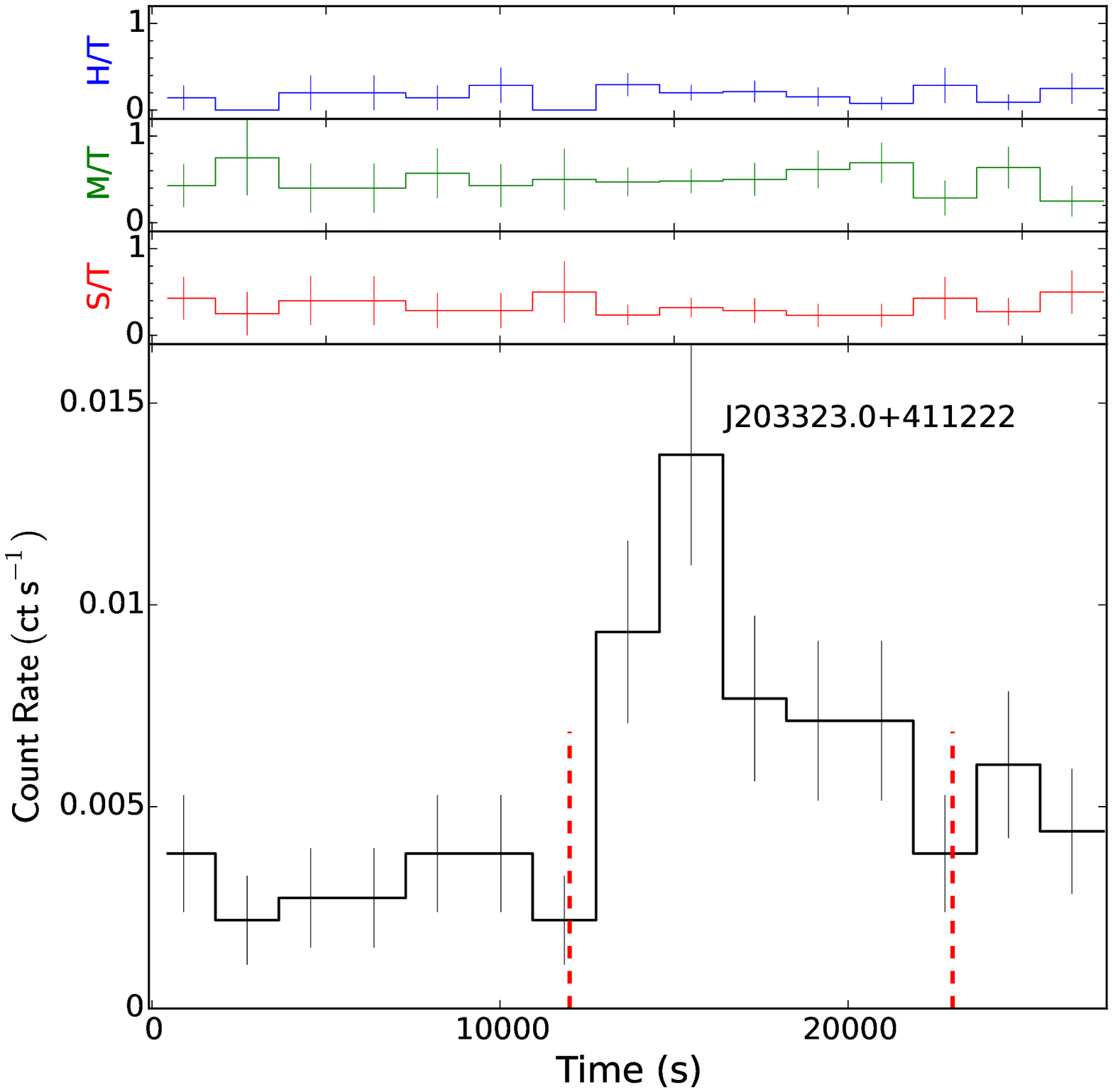}
\includegraphics[width=0.33\textwidth]{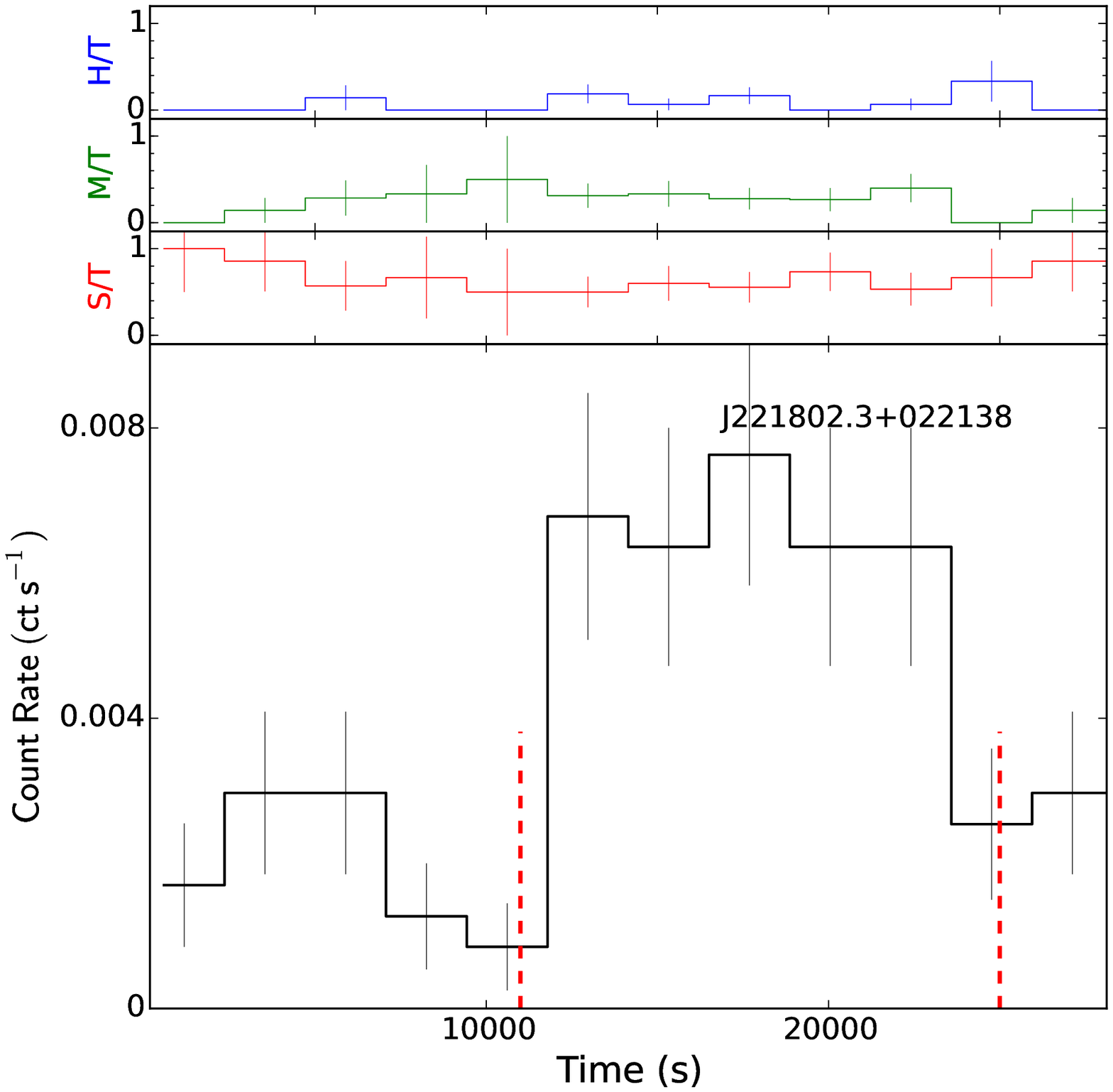}
\includegraphics[width=0.33\textwidth]{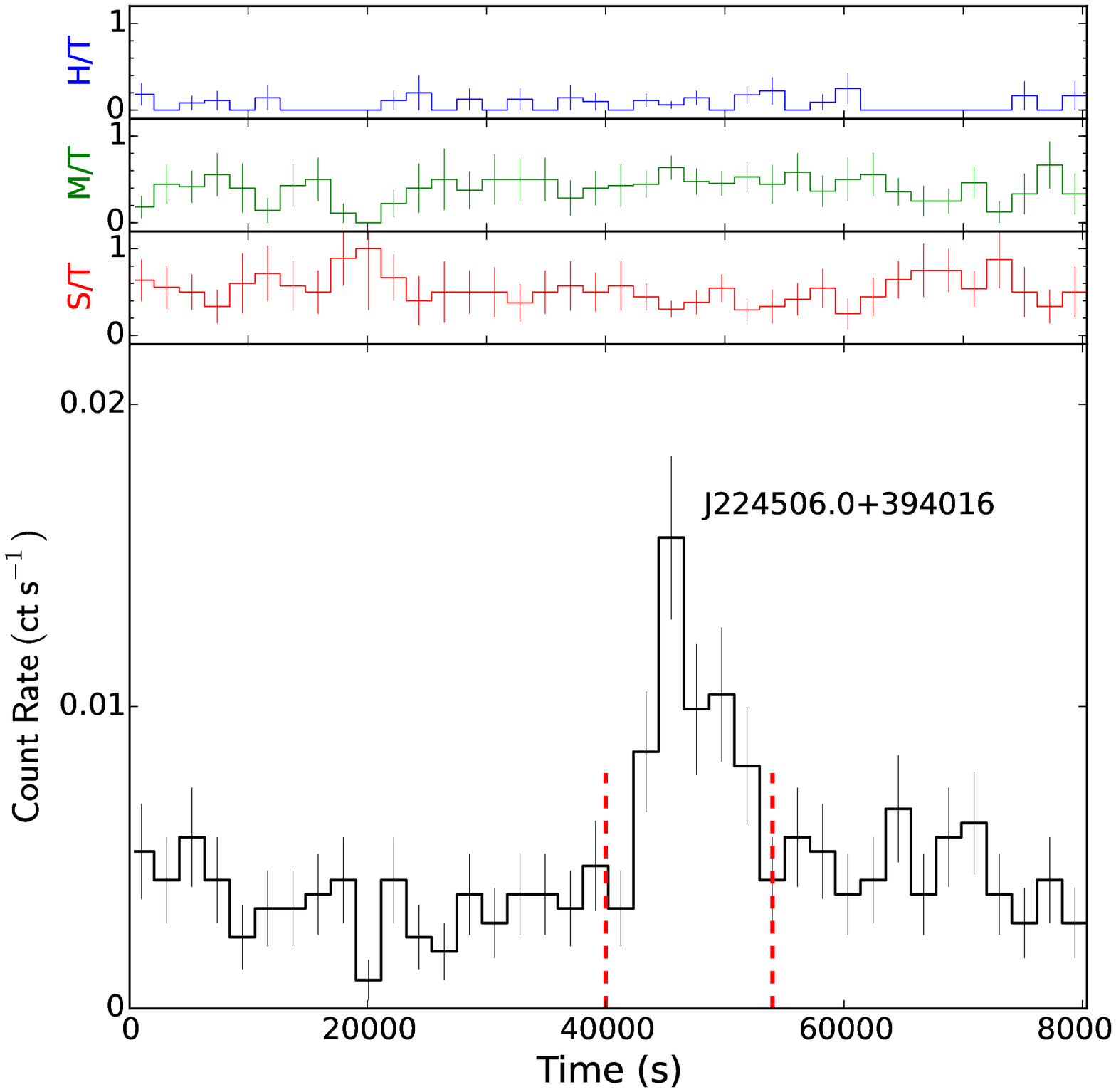}
\caption[]{Stars with X-ray flares.
The red dashed lines show the beginning times and/or the ending times of the flares, which are roughly determined by eyes.}
\label{flare.fig}
\end{figure*}

We selected stars with X-ray flares by eye (Table \ref{flare.tab}).
%
The light curves of the flares are clearly variable (Figure \ref{flare.fig}).
Some flares are like compact solar flares with short duration (e.g., J122837.2+015720);
some flares show quite long duration (e.g., J055207.8+322639);
some flares have a low rise shape (e.g., J063324.2+222837);
some flares have strong absorption (e.g., J193015.7+493209).
The morphological differences indicate different processes of energy release in these flares \citep{Pandey2008}.
For example, compact flares suggest a quick energy release,
while a prolonged energy release is preferred for long decay flares \citep{Pallavicini1977,Pallavicini1988}.

\begin{table}
\begin{center}
\begin{threeparttable}
\caption[]{Stars with X-ray flares in our sample.}
\label{flare.tab}
\begin{tabular}{lccc}
\hline\noalign{\smallskip}
Object  & Subclass &  ObsID  & Duration       \\
        &          &         &   (s)              \\
  (1)   &     (2)  &  (3)    &   (4)             \\
\hline\noalign{\tiny}
J004118.6+405159 &  G3 & 2049 &   4500      \\
J055207.8+322639 & G2 & 13656 & 43000 \\
J063214.1+045627 &  A2IV & 3750 & $>$15000  \\
J063324.2+222837 &  M3 & 4467  & $>$20000  \\
J122837.2+015720 & M0 & 1712 & 4000 \\
J131156.7-011311 &  K1 & 1663  & 4000      \\
J151628.4+070241 & K5 & 5807 & 15000 \\
J155710.7+055321 & K3 & 11363 & $>$5000     \\
J193015.7+493209 &  G9  & 13612 & 6000      \\
J203323.0+411222 &  K0  & 10956 & 11000     \\
J221802.3+022138 &  K1  & 2981 & 14000     \\
J224506.0+394016 & K4 & 2195  &14000 \\
\noalign{\smallskip}\hline
\end{tabular}
\end{threeparttable}
\end{center}
\end{table}

\section{Conclusion}
\label{summary.sec}

The $Chandra$ and LAMOST data allow us to probe stellar X-ray activity over a wide range of stellar parameters.
In this paper, we cross-matched the $Chandra$ point source catalog \citep{Wang2016b} and the LAMOST database (DR4),
and obtained a sample of 1086 stars with X-ray emission and at least one LAMOST spectral observation.
Finally, the X-ray-to-bolometric luminosity $R_X$ was estimated for 484 sources,
for which the optical magnitude and complete stellar parameters ($T_{\rm eff}$, log$g$, $E(B-V)$, distance)
can be obtained.
Using this sample, we carefully studied the correlation between stellar X-ray activity and stellar parameters.
We are aware that our sample is not unbiased, because the X-ray data come from targeted observations. Therefore, our results must not be taken as representative of the whole Galactic stellar population, but rather as a hint for further investigations based on larger samples.

The main conclusions of this paper are summarized as follows:

(1) The $R_X$ distributions of G and K stars show a bimodal profile.
Although our sample is not inhomogeneous, it is consistent with previous studies finding bimodal stellar activity.
For G stars, there are more inactive stars than active ones, while for K stars, more active stars are apparent.
This bimodality represents two sub-populations with different coronal temperatures.
Stars with a hotter corona --- observationally with a higher hardness ratio --- have a higher X-ray $R_X$ value.
Using metallicity and velocity information, we ascertain that most of those G and K stars (both the active and inactive ones) are located in the thin disk. Hence, the observed bimodality cannot be associated to different sub-populations in younger and older structures of the Galaxy.

(2) There are $\approx$ 51 giants showing X-ray emission, and some have high X-ray activity.
This confirms previous observations of stellar activity
(e.g., X-ray emission, chromospheric emission, photometric variability) among giants.
However, one should note that
some giants or sub-giants showing stellar activity may be unrecognized binary systems \citep{Ozdarcan2018}.
Also, some A-type stars show X-ray activity.
Our data are not sufficient to associate the X-ray emission to the A-type stars themselves
or their unresolved low-mass companions \citep{Schroder2007}.
These stars may be interesting candidates for follow-up observations.

(3) We find no trend of $R_X$ with stellar age.
However, owing to the small sample and the uncertainty on our ages, we cannot confirm whether
the activity still decreases with age after 2 Gyr or keeps constant for older stars.
A trough with the lowest $R_X$ values around 2 Gyr seems consistent with \citet{Pace2013},
but the reason for this dip is not clear.

(4) We calculated the H$\alpha$ emission index $R_{\rm H\alpha}$ using the EWs from LAMOST spectra.
The proxy of coronal activity ($R_X$) and the proxy of chromospheric activity ($R_{\rm H\alpha}$)
show a positive tight correlation:
${\rm log}~R_X = (1.16 \pm 0.28)\times~{\rm log}~R_{\rm H\alpha} + (1.22 \pm 1.03)$.

(5) We studied the light-curve morphology of the flares for twelve stars.
The light curves show evident morphological differences (e.g., duration time and shape),
indicating different processes of energy release, and possibly different coronal structures.
%
We are currently planning a follow-up study to search for stellar X-ray flares using the {\it Chandra} archive,
and investigate the flare properties in different stellar types.

\begin{acknowledgements}
We especially thank the anonymous referee for his/her thorough report and helpful comments
and suggestions that have significantly improved the paper.
This work has made use of data obtained from the {\it Chandra} Data Archive, and software provided by the {\it Chandra} X-ray Center (CXC) in the application packages CIAO.
Guoshoujing Telescope (the Large Sky Area Multi-Object Fiber Spectroscopic Telescope LAMOST) is a National Major Scientific Project built by the Chinese Academy of Sciences. Funding for the project has been provided by the National Development and Reform Commission. LAMOST is operated and managed by the National Astronomical Observatories, Chinese Academy of Sciences.
We acknowledge use of the SIMBAD database and the VizieR catalogue access tool, operated at CDS, Strasbourg, France, and of Astropy, a community-developed core Python package for Astronomy (Astropy Collaboration, 2013). We are grateful for support from the National Science Foundation of China (NSFC, Nos. 11273028, 11333004, 11603035, 11603038, and 11503054). RS acknowledges support from a Curtin University Senior Research Fellowship; he is also grateful for support, discussions and hospitality at the Strasbourg Observatory during part of this work.
\end{acknowledgements}


\begin{thebibliography}



\bibitem[Ag{\"u}eros et al.(2009)]{Agueros2009} Ag{\"u}eros, M.~A.,
Anderson, S.~F., Covey, K.~R., et al.\ 2009, \apjs, 181, 444




\bibitem[Audard et al.(2000)]{Audard2000} Audard, M., G{\"u}del, M., Drake,
J.~J., \& Kashyap, V.~L.\ 2000, \apj, 541, 396

\bibitem[Auri{\`e}re et al.(2015)]{Auriere2015} Auri{\`e}re, M., Konstantinova-Antova, R., Charbonnel, C., et al.\ 2015, \aap, 574, A90



\bibitem[Baliunas et al.(1995)]{Baliunas1995} Baliunas, S.~L., Donahue,
R.~A., Soon, W.~H., et al.\ 1995, \apj, 438, 269


\bibitem[Baliunas \& Jastrow(1990)]{Baliunas1990} Baliunas, S., \& Jastrow, R.\ 1990, \nat, 348, 520



\bibitem[Benz \& G{\"u}del(2010)]{Benz2010} Benz, A.~O., \& G{\"u}del, M.\ 2010, \araa, 48, 241

\bibitem[Bertone et al.(2004)]{Bertone2004} Bertone, E., Buzzoni, A., Ch{\'a}vez, M., \& Rodr{\'{\i}}guez-Merino, L.~H.\ 2004, \aj, 128, 829





\bibitem[Blackman \& Thomas(2015)]{Blackman2015} Blackman, E.~G., \& Thomas, J.~H.\ 2015, \mnras, 446, L51


\bibitem[Booth et al.(2017)]{Booth2017} Booth, R.~S., Poppenhaeger, K.,
Watson, C.~A., Silva Aguirre, V., \& Wolk, S.~J.\ 2017, \mnras, 471, 1012





\bibitem[Cardini \& Cassatella(2007)]{Cardini2007} Cardini, D., \& Cassatella, A.\ 2007, \apj, 666, 393

\bibitem[Carroll \& Ostlie(1996)]{Carroll1996} Carroll, B. W., \& Ostlie, D. A. 1996,
An Introduction to Modern Astrophysics (Cambridge: Pearson)

\bibitem[Castelli \& Kurucz(2004)]{ck04} Castelli, F., \& Kurucz, R.~L.\ 2004, arXiv:astro-ph/0405087


\bibitem[Charbonneau(2010)]{Charbonneau2010} Charbonneau, P.\ 2010, Living
Reviews in Solar Physics, 7, 3


\bibitem[Ciardi et al.(2011)]{Ciardi2011} Ciardi, D.~R., von Braun, K.,
Bryden, G., et al.\ 2011, \aj, 141, 108


\bibitem[Cincunegui et al.(2007)]{Cincunegui2007} Cincunegui, C.,
D{\'{\i}}az, R.~F., \& Mauas, P.~J.~D.\ 2007, \aap, 469, 309


\bibitem[Cranmer \& Saar(2011)]{Cranmer2011} Cranmer, S.~R., \& Saar, S.~H.\ 2011, \apj, 741, 54


\bibitem[Cranmer et al.(2007)]{Cranmer2007} Cranmer, S.~R., van
Ballegooijen, A.~A., \& Edgar, R.~J.\ 2007, \apjs, 171, 520


\bibitem[Cui et al.(2012)]{Cui2012} Cui, X.-Q., Zhao, Y.-H., Chu, Y.-Q., et
al.\ 2012, Research in Astronomy and Astrophysics, 12, 1197


\bibitem[Cully et al.(1994)]{Cully1994} Cully, S.~L., Fisher, G.~H.,
Abbott, M.~J., \& Siegmund, O.~H.~W.\ 1994, \apj, 435, 449


\bibitem[Cully et al.(1993)]{Cully1993} Cully, S.~L., Siegmund, O.~H.~W.,
Vedder, P.~W., \& Vallerga, J.~V.\ 1993, \apjl, 414, L49


\bibitem[Di Stefano \& Kong(2003)]{Di Stefano2003} Di Stefano, R., \& Kong, A.~K.~H.\ 2003, arXiv:astro-ph/0311374

\bibitem[Donahue(1993)]{Donahue1993} Donahue, R. A. 1993, Ph.D. Thesis, New Mexico State Univ.

\bibitem[Doyle \& Butler(1985)]{Doyle1985} Doyle, J.~G., \& Butler, C.~J.\ 1985, \nat, 313, 378



  \bibitem[Fang et al.(2018)]{Fang2018} Fang, X.-S., Zhao, G., Zhao, J.-K., \& Bharat Kumar, Y.\ 2018, \mnras, 476, 908

\bibitem[Favata et al.(2000)]{Favata2000} Favata, F., Micela, G.,
\& Reale, F.\ 2000, \aap, 354, 1021

\bibitem[Feigelson et al.(2004)]{Feigelson2004} Feigelson, E.~D.,
Hornschemeier, A.~E., Micela, G., et al.\ 2004, \apj, 611, 1107









\bibitem[Flaccomio et al.(2003)]{Flaccomio2003} Flaccomio, E., Damiani, F.,
Micela, G., et al.\ 2003, \apj, 582, 398

\bibitem[Flewelling et al.(2016)]{Flewelling2016} Flewelling, H.~A., Magnier, E.~A., Chambers, K.~C., et al.\ 2016, arXiv:1612.05243

\bibitem[Gallet et al.(2017)]{Gallet2017} Gallet, F., Charbonnel, C., 
Amard, L., et al.\ 2017, \aap, 597, A14 

\bibitem[G{\"u}del(2004)]{Gudel2004} G{\"u}del, M.\ 2004, \aapr, 12, 71


\bibitem[G{\"u}del(2002)]{Gudel2002a} G{\"u}del, M.\ 2002, \araa, 40, 217

\bibitem[G{\"u}del et al.(2002)]{Gudel2002} G{\"u}del, M., Audard, M.,
Skinner, S.~L., \& Horvath, M.~I.\ 2002, \apjl, 580, L73

\bibitem[G{\"u}del et al.(2014)]{Gudel2014} G{\"u}del, M., Dvorak, R., 
Erkaev, N., et al.\ 2014, Protostars and Planets VI, 883 



\bibitem[Getman et al.(2008)]{Getman2008} Getman, K.~V., Feigelson, E.~D.,
Micela, G., et al.\ 2008, \apj, 688, 437-455





\bibitem[Golub et al.(1983)]{Golub1983} Golub, L., Harnden, F.~R., Jr., Maxson, C.~W., et al.\ 1983, \apj, 271, 264


\bibitem[Gondoin(2012)]{Gondoin2012} Gondoin, P.\ 2012, \aap, 546, A117


\bibitem[Graffagnino et al.(1995)]{Graffagnino1995} Graffagnino, V.~G.,
Wonnacott, D., \& Schaeidt, S.\ 1995, \mnras, 275, 129




\bibitem[Gray et al.(2003)]{Gray2003} Gray, R.~O., Corbally, C.~J.,
Garrison, R.~F., McFadden, M.~T., \& Robinson, P.~E.\ 2003, \aj, 126, 2048




\bibitem[Guo et al.(2016)]{Guo2016} Guo, J.-C., Liu, C.,
\& Liu, J.-F.\ 2016, Research in Astronomy and Astrophysics, 16, 44


\bibitem[Haisch et al.(1987)]{Haisch1987} Haisch, B.~M., Butler, C.~J.,
Doyle, J.~G., \& Rodono, M.\ 1987, \aap, 181, 96









\bibitem[Henry et al.(1996)]{Henry1996} Henry, T.~J., Soderblom, D.~R.,
Donahue, R.~A., \& Baliunas, S.~L.\ 1996, \aj, 111, 439













\bibitem[Jeffries(2014)]{Jeffries2014} Jeffries, R.~D.\ 2014, EAS
Publications Series, 65, 289

\bibitem[Jeffries et al.(2006)]{Jeffries2006} Jeffries, R.~D., Evans,
P.~A., Pye, J.~P., \& Briggs, K.~R.\ 2006, \mnras, 367, 781


\bibitem[Jenkins et al.(2008)]{Jenkins2008} Jenkins, J.~S., Jones,
H.~R.~A., Pavlenko, Y., et al.\ 2008, \aap, 485, 571


\bibitem[Jenkins et al.(2006)]{Jenkins2006} Jenkins, J.~S., Jones,
H.~R.~A., Tinney, C.~G., et al.\ 2006, \mnras, 372, 163


\bibitem[Jenkins et al.(2011)]{Jenkins2011} Jenkins, J.~S., Murgas, F.,
Rojo, P., et al.\ 2011, \aap, 531, A8


\bibitem[Jester et al.(2005)]{Jester2005} Jester, S., Schneider, D.~P.,
Richards, G.~T., et al.\ 2005, \aj, 130, 873

\bibitem[Judge et al.(2003)]{Judge2003} Judge, P.~G., Solomon, S.~C.,
\& Ayres, T.~R.\ 2003, \apj, 593, 534

\bibitem[Katsova \& Livshits(2011)]{Katsova2011} Katsova, M.~M., \& Livshits, M.~A.\ 2011, Astronomy Reports, 55, 1123

\bibitem[Katsova et al.(2016)]{Katsova2016} Katsova, M.~M., Livshits,
M.~A., Mishenina, T.~V.,
\& Nizamov, B.~A.\ 2016, 19th Cambridge Workshop on Cool Stars, Stellar Systems, and the Sun (CS19), 124


\bibitem[Klimchuk(2006)]{Klimchuk2006} Klimchuk, J.~A.\ 2006, \solphys,
234, 41


\bibitem[Kraft(1967)]{Kraft1967} Kraft, R.~P.\ 1967, \apj, 150, 551




\bibitem[Kuerster \& Schmitt(1996)]{Kuerster1996} Kuerster, M., \& Schmitt, J.~H.~M.~M.\ 1996, \aap, 311, 211




\bibitem[Lachaume et al.(1999)]{Lachaume1999} Lachaume, R., Dominik, C.,
Lanz, T., \& Habing, H.~J.\ 1999, \aap, 348, 897






\bibitem[Lucy \& White(1980)]{Lucy1980} Lucy, L.~B., \& White, R.~L.\ 1980, \apj, 241, 300


\bibitem[Luo et al.(2015)]{Luo2015} Luo, A.-L., Zhao, Y.-H., Zhao, G., et
al.\ 2015, Research in Astronomy and Astrophysics, 15, 1095

\bibitem[Lyra \& Porto de Mello(2005)]{Lyra2005} Lyra, W., \& Porto de Mello, G.~F.\ 2005, \aap, 431, 329


\bibitem[Maccacaro et al.(1988)]{Maccacaro1988} Maccacaro, T., Gioia,
I.~M., Wolter, A., Zamorani, G., \& Stocke, J.~T.\ 1988, \apj, 326, 680


\bibitem[Mamajek \& Hillenbrand(2008)]{Mamajek2008} Mamajek, E.~E., \& Hillenbrand, L.~A.\ 2008, \apj, 687, 1264-1293



\bibitem[Mart{\'{\i}}nez-Arn{\'a}iz et al.(2011)]{Martinez-Arnaiz2011}
Mart{\'{\i}}nez-Arn{\'a}iz, R., L{\'o}pez-Santiago, J., Crespo-Chac{\'o}n,
I., \& Montes, D.\ 2011, \mnras, 414, 2629



\bibitem[Masseron \& Gilmore(2015)]{Masseron2015} Masseron, T., \& Gilmore, G.\ 2015, \mnras, 453, 1855

\bibitem[Mathioudakis \& Doyle(1989)]{Mathioudakis1989} Mathioudakis, M., \& Doyle, J.~G.\ 1989, \aap, 224, 179

\bibitem[Mathur et al.(2014)]{Mathur2014} Mathur, S., Garc{\'{\i}}a, R.~A.,
Ballot, J., et al.\ 2014, \aap, 562, A124




\bibitem[Micela et al.(1985)]{Micela1985} Micela, G., Sciortino, S., Serio,
S., et al.\ 1985, \apj, 292, 172


\bibitem[Narain \& Ulmschneider(1990)]{Narain1990} Narain, U., \& Ulmschneider, P.\ 1990, \ssr, 54, 377



\bibitem[Ol{\'a}h \& Strassmeier(2002)]{Olah2002} Ol{\'a}h, K., \& Strassmeier, K.~G.\ 2002, Astronomische Nachrichten, 323, 361


\bibitem[Osten \& Brown(1999)]{Osten1999} Osten, R.~A., \& Brown, A.\ 1999, \apj, 515, 746

\bibitem[Osten et al.(2007)]{Osten2007} Osten, R.~A., Drake, S., Tueller,
  J., et al.\ 2007, \apj, 654, 1052


\bibitem[{\"O}zdarcan \& Dal(2018)]{Ozdarcan2018} {\"O}zdarcan, O., \& Dal, H.~A.\ 2018, arXiv:1801.06087


\bibitem[Pace(2013)]{Pace2013} Pace, G.\ 2013, \aap, 551, L8


\bibitem[Pace \& Pasquini(2004)]{Pace2004} Pace, G., \& Pasquini, L.\ 2004, \aap, 426, 1021


\bibitem[Pallavicini et al.(1981)]{Pallavicini1981} Pallavicini, R., Golub,
L., Rosner, R., et al.\ 1981, \apj, 248, 279

\bibitem[Pallavicini et al.(1988)]{Pallavicini1988} Pallavicini, R.,
Monsignori-Fossi, B.~C., Landini, M.,
\& Schmitt, J.~H.~M.~M.\ 1988, \aap, 191, 109


\bibitem[Pallavicini et al.(1977)]{Pallavicini1977} Pallavicini, R., Serio,
S., \& Vaiana, G.~S.\ 1977, \apj, 216, 108

\bibitem[Pallavicini et al.(1990)]{Pallavicini1990} Pallavicini, R.,
Tagliaferri, G., \& Stella, L.\ 1990, \aap, 228, 403

\bibitem[Pandey \& Singh(2008)]{Pandey2008} Pandey, J.~C., \& Singh, K.~P.\ 2008, \mnras, 387, 1627

\bibitem[Panzera et al.(1999)]{Panzera1999} Panzera, M.~R., Tagliaferri, G., Pasinetti, L., \& Antonello, E.\ 1999, \aap, 348, 161


\bibitem[Parkin et al.(2009)]{Parkin2009} Parkin, E.~R., Pittard, J.~M.,
Hoare, M.~G., Wright, N.~J., \& Drake, J.~J.\ 2009, \mnras, 400, 629

\bibitem[Pease et al.(2006)]{Pease2006} Pease, D.~O., Drake, J.~J., \& Kashyap, V.~L.\ 2006, \apj, 636, 426


\bibitem[Peres et al.(2000)]{Peres2000} Peres, G., Orlando, S., Reale, F.,
Rosner, R., \& Hudson, H.\ 2000, \apj, 528, 537


\bibitem[Pizzolato et al.(2003)]{Pizzolato2003} Pizzolato, N., Maggio, A.,
Micela, G., Sciortino, S., \& Ventura, P.\ 2003, \aap, 397, 147







\bibitem[Pye et al.(2015)]{Pye2015} Pye, J.~P., Rosen, S., Fyfe, D.,
\& Schr{\"o}der, A.~C.\ 2015, \aap, 581, A28



\bibitem[Reale et al.(2004)]{Reale2004} Reale, F., G{\"u}del, M., Peres,
G., \& Audard, M.\ 2004, \aap, 416, 733






\bibitem[Reiners et al.(2014)]{Reiners2014} Reiners, A., Sch{\"u}ssler, M.,
\& Passegger, V.~M.\ 2014, \apj, 794, 144



\bibitem[Robertson et al.(2015)]{Robertson2015} Robertson, P., Roy, A., 
\& Mahadevan, S.\ 2015, \apjl, 805, L22 

\bibitem[Rosner et al.(1985)]{Rosner1985} Rosner, R., Golub, L.,
\& Vaiana, G.~S.\ 1985, \araa, 23, 413




\bibitem[Saar et al.(1998)]{Saar1998} Saar, S.~H., Butler, R.~P.,
\& Marcy, G.~W.\ 1998, \apjl, 498, L153





\bibitem[Schmitt et al.(1990)]{Schmitt1990} Schmitt, J.~H.~M.~M., Collura,
A., Sciortino, S., et al.\ 1990, \apj, 365, 704

\bibitem[Schmitt \& Favata(1999)]{Schmitt1999} Schmitt, J.~H.~M.~M., \& Favata, F.\ 1999, \nat, 401, 44


\bibitem[Schmitt et al.(1985)]{Schmitt1985} Schmitt, J.~H.~M.~M., Golub,
L., Harnden, F.~R., Jr., et al.\ 1985, \apj, 290, 307



\bibitem[Schr{\"o}der \& Schmitt(2007)]{Schroder2007} Schr{\"o}der, C., \& Schmitt, J.~H.~M.~M.\ 2007, \aap, 475, 677






\bibitem[Schrijver et al.(1992)]{Schrijver1992} Schrijver, C.~J., Dobson,
A.~K., \& Radick, R.~R.\ 1992, \aap, 258, 432

\bibitem[Simon \& Drake(1989)]{Simon1989} Simon, T., \& Drake, S.~A.\ 1989, \apj, 346, 303


\bibitem[Simon et al.(1985)]{Simon1985} Simon, T., Herbig, G.,
\& Boesgaard, A.~M.\ 1985, \apj, 293, 551


\bibitem[Sissa et al.(2016)]{Sissa2016} Sissa, E., Gratton, R., Desidera,
S., et al.\ 2016, \aap, 596, A76


\bibitem[Skumanich(1985)]{Skumanich1985} Skumanich, A.\ 1985, Australian
Journal of Physics, 38, 971


\bibitem[Skumanich(1972)]{Skumanich1972} Skumanich, A.\ 1972, \apj, 171,
565

\bibitem[Soderblom(2010)]{Soderblom2010} Soderblom, D.~R.\ 2010, \araa, 48,
581

\bibitem[Soderblom et al.(1991)]{Soderblom1991} Soderblom, D.~R., Duncan,
D.~K., \& Johnson, D.~R.~H.\ 1991, \apj, 375, 722








\bibitem[Stelzer et al.(2012)]{Stelzer2012} Stelzer, B., Alcal{\'a}, J.,
Biazzo, K., et al.\ 2012, \aap, 537, A94


\bibitem[Stelzer et al.(2013)]{Stelzer2013} Stelzer, B., Marino, A.,
Micela, G., L{\'o}pez-Santiago, J., \& Liefke, C.\ 2013, \mnras, 431, 2063


\bibitem[Sterzik \& Schmitt(1997)]{Sterzik1997} Sterzik, M.~F., \& Schmitt, J.~H.~M.~M.\ 1997, \aj, 114, 1673

\bibitem[Stocke et al.(1983)]{Stocke1983} Stocke, J.~T., Liebert, J.,
Gioia, I.~M., et al.\ 1983, \apj, 273, 458


\bibitem[Stocke et al.(1991)]{Stocke1991} Stocke, J.~T., Morris, S.~L.,
Gioia, I.~M., et al.\ 1991, \apjs, 76, 813




\bibitem[Tagliaferri et al.(1991)]{Tagliaferri1991} Tagliaferri, G., White,
N.~E., Doyle, J.~G., et al.\ 1991, \aap, 251, 161


\bibitem[Telleschi et al.(2007)]{Telleschi2007} Telleschi, A., G{\"u}del,
M., Briggs, K.~R., Audard, M., \& Palla, F.\ 2007, \aap, 468, 425

\bibitem[Testa et al.(2015)]{Testa2015} Testa, P., Saar, S.~H.,
\& Drake, J.~J.\ 2015, Philosophical Transactions of the Royal Society of London Series A, 373, 20140259




\bibitem[Vaiana(1983)]{Vaiana1983} Vaiana, G.~S.\ 1983, Solar and Stellar
Magnetic Fields: Origins and Coronal Effects, 102, 165

\bibitem[Vaiana et al.(1981)]{Vaiana1981} Vaiana, G.~S., Cassinelli, J.~P.,
Fabbiano, G., et al.\ 1981, \apj, 245, 163

\bibitem[Vilhu(1984)]{Vilhu1984} Vilhu, O.\ 1984, \aap, 133, 117


\bibitem[Vollmer et al.(2010)]{Vollmer2010} Vollmer, B., Gassmann, B.,
Derri{\`e}re, S., et al.\ 2010, \aap, 511, A53

\bibitem[Wang et al.(2016a)]{Wang2016a} Wang, S., Liu, J., Qiu, Y., et al.\
2016a, \apjs, 224, 40

\bibitem[Wang et al.(2016b)]{Wang2016b} Wang, S., Qiu, Y., Liu, J.,
\& Bregman, J.~N.\ 2016b, \apj, 829, 20

\bibitem[Wang et al.(2016)]{Wang2016} Wang, J., Shi, J., Zhao, Y., et al.\
2016, \mnras, 456, 672









\bibitem[Wilson(1963)]{Wilson1963} Wilson, O.~C.\ 1963, \apj, 138, 832


\bibitem[Wright(2004)]{Wright2004} Wright, J.~T.\ 2004, \aj, 128, 1273


\bibitem[Wright et al.(2010)]{Wright2010} Wright, N.~J., Drake, J.~J.,
\& Civano, F.\ 2010, \apj, 725, 480


\bibitem[Wright et al.(2011)]{Wright2011} Wright, N.~J., Drake, J.~J.,
Mamajek, E.~E., \& Henry, G.~W.\ 2011, \apj, 743, 48


\bibitem[Wu et al.(2011)]{Wu2011} Wu, Y., Luo, A.-L., Li, H.-N., et al.\
2011, Research in Astronomy and Astrophysics, 11, 924


\bibitem[Xiang et al.(2017a)]{Xiang2017a} Xiang, M., Liu, X., Shi, J., et
al.\ 2017a, \apjs, 232, 2


\bibitem[Xiang et al.(2017b)]{Xiang2017b} Xiang, M.-S., Liu, X.-W., Yuan,
H.-B., et al.\ 2017b, \mnras, 467, 1890


\bibitem[Yang et al.(2017)]{Yang2017} Yang, H., Liu, J., Gao, Q., et al.\
2017, \apj, 849, 36






\bibitem[Zacharias et al.(2013)]{Zacharias2013} Zacharias, N., Finch, C.~T., Girard, T.~M., et al.\ 2013, \aj, 145, 44

\bibitem[{\v Z}erjal et al.(2017)]{Zerjal2017} {\v Z}erjal, M., Zwitter,
T., Matijevi{\v c}, G., et al.\ 2017, \apj, 835, 61



\bibitem[Zhao et al.(2013)]{Zhao2013} Zhao, J.~K., Oswalt, T.~D., Zhao, G.,
et al.\ 2013, \aj, 145, 140

\bibitem[Zhao et al.(2012)]{Zhao2012} Zhao, G., Zhao, Y.-H., Chu, Y.-Q.,
Jing, Y.-P.,
\& Deng, L.-C.\ 2012, Research in Astronomy and Astrophysics, 12, 723


\bibitem[Zhu et al.(2017)]{Zhu2017} Zhu, H., Tian, W., Li, A.,
\& Zhang, M.\ 2017, \mnras, 471, 3494


\bibitem[Zinnecker \& Preibisch(1994)]{Zinnecker1994} Zinnecker, H., \& Preibisch, T.\ 1994, \aap, 292, 152




\end{thebibliography}
\end{document}